\documentclass{jfm_draft}

\usepackage{newtxtext}
\usepackage{newtxmath}
\usepackage{hyperref}
\hypersetup{
    colorlinks = true,
    urlcolor   = blue,
    citecolor  = black,
}
\newcommand{\RomanNumeralCaps}[1]
\linenumbers

\usepackage[utf8]{inputenc}
\usepackage{graphicx}
\usepackage{overpic}
\usepackage[sort,numbers]{natbib}
\usepackage{mathtools}
\usepackage{amssymb}
\usepackage{amsbsy}
\usepackage{bm}
\usepackage{color}
\usepackage{tikz}
\usepackage[cal=euler]{mathalpha}
\usepackage{xcolor} 
\usepackage[nameinlink,noabbrev]{cleveref}

\usepackage{my_macros}

\allowdisplaybreaks

\title{Rapidly yawing spheroids in viscous shear flow: Emergent loss of symmetry}

\date{\today}

\shorttitle{Rapidly yawing spheroids generate asymmetric emergent effects}
\shortauthor{M. P. Dalwadi}
\author{M. P. Dalwadi\aff{1,2\corresp{\email{dalwadi@maths.ox.ac.uk}}}}

\affiliation{\aff{1}Mathematical Institute, University of Oxford, Oxford OX2 6GG, UK \\
\aff{2}Department of Mathematics, University College London, London WC1H 0AY, UK}

\date{}

\begin{document}

\maketitle

\begin{abstract}
We investigate the emergent three-dimensional (3D) dynamics of a rapidly yawing spheroidal swimmer interacting with a viscous shear flow. We show that the rapid yawing generates non-axisymmetric emergent effects, with the active swimmer behaving as an effective passive particle with two orthogonal planes of symmetry. We also demonstrate that this effective asymmetry generated by the rapid yawing can cause chaotic behaviour in the emergent dynamics, in stark contrast to the emergent dynamics generated by rapidly rotating spheroids, which are equivalent to those of effective passive spheroids. In general, we find that the shape of the equivalent effective particle under rapid yawing is different to the average shape of the active particle. Moreover, despite having two planes of symmetry, the equivalent passive particle is not an ellipsoid in general, except for specific scenarios in which the effective shape is a spheroid. In these scenarios, we calculate analytically the equivalent aspect ratio of the effective spheroid. We use a multiple scales analysis for systems to derive the emergent swimmer behaviour, which requires solving a nonautonomous nonlinear 3D dynamical system, and we validate our analysis via comparison to numerical simulations.
\end{abstract}

\section{Introduction}

Bulk properties of particle suspensions depend on particle orientation, and particle-particle interactions can be neglected for sufficiently dilute suspensions \citep{leal1971effect,saintillan2015theory}. Hence, in many cases interaction with the local flow is a key factor in particle orientation. For small enough particles, viscous effects dominate and orientation is mainly forced by the local shear flow approximation. Thus, the rotational dynamics of single particles in viscous shear flows are of fundamental interest in fluid mechanics.

A classic result in fluid mechanics is that passive spheroids undergo non-uniform rotation in viscous shear flow. Their angular dynamics are governed by Jeffery's equations \citep{jeffery1922motion,taylor1923motion}, and their periodic but uneven rotations are called Jeffery's orbits. The precise nature of the Jeffery's orbit depends on the spheroid aspect ratio $\ratio \in (0, \infty)$ via the quantity $(\ratio^2 - 1)/(\ratio^2 + 1)$, with values of $\ratio$ further from unity causing more uneven dynamics.

More generally, Jeffery's equations hold for a wider class of particles beyond spheroids, including passive axisymmetric objects \citep{bretherton1962motion,brenner1964stokes}, parameterised via a coefficient called the Bretherton parameter, $\Breth$. The derived governing equations for axisymmetric objects are equivalent to Jeffery's equations when equating $\Breth = (\ratio^2 - 1)/(\ratio^2 + 1)$. Therefore, axisymmetric objects with $\Breth \in (-1,1)$ demonstrate angular dynamics in shear flow equivalent to those of a spheroid with effective aspect ratio.

Asymmetry of particles can induce fundamentally different behaviours to axisymmetric objects \citep{hinch1979rotation,roggeveen2022motion,miara2024dynamics}. For example, helicoidal objects are governed by modified versions of Jeffery's equations, with extra terms characterised by two additional coefficients that account for chiral effects \citep{ishimoto2020helicoidal}. In addition, the loss of axial symmetry caused by replacing spheroids with triaxial ellipsoids, and more generally to particles with two orthogonal planes of symmetry, can generate chaotic dynamics \citep{yarin1997chaotic,thorp2019motion}.

In the studies mentioned in the paragraph above, the particles are passive. However, particle activity makes interactions with fluid flow much more complicated \citep{junot2019swimming,lauga2009hydrodynamics,saintillan2018rheology,elgeti2015physics,wittkowski2012self}. Recent work deriving the emergent behaviour of single active particles in shear flow has shown that self-propelled objects exhibiting fast-scale periodic motion can generate emergent slow angular dynamics in shear flow \citep{ishimoto2023jeffery}. For example, oscillatory yawing of ellipses (2D) \citep{walker2022effects}, and constant rotation of spheroids and helicoidal objects (3D) \citep{dalwadi2024generalisedparti,dalwadi2024generalisedpartii}. These studies use the method of multiple scales \citep{hinch_1991} to understand the nonlinear interaction between the fast self-propulsion and slow shear flow, and demonstrate that this generates emergent angular dynamics equivalent to those of a passive particle. The calculated shapes of these equivalent effective particles depend on the type of fast motion and the original shapes. Notably, in these scenarios the effective shape maintains the hydrodynamic symmetries of the original particle. The method of multiple scales has also been used recently to understand the effective dynamics of particles in unsteady flow fields \citep{pujara2023wave,ventrella2023microswimmer,ma2022reaching}.

The specific type of activity we are interested in here is rapid yawing in 3D. Undulatory motion can be observed in many microswimmers, especially flagellates \citep{guasto2012fluid}, for example \emph{Chlamydomonas} \citep{leptos2023phototaxis} and spermatozoon \citep{shaebani2020computational}. The latter is particularly well studied due to the implications for understanding sperm motility, impacting understanding of motility-based male fertility diagnostics, reproductive toxicology and basic sperm function \citep{walker2020computer,gaffney2011mammalian}.

For simplicity and analytic tractability, we neglect the complexities of how exactly the rapid motion arises at the microswimmer scale. With the goal of gaining insight into how rapid microscale motion interacts with a far-field shear flow, we consider a simple model of self-generated rapid yawing of a rigid spheroid in steady Stokes flow, with an accompanying self-generated translation. Some care must be taken in considering the limit of rapid yawing while using steady Stokes flow, since very fast oscillations can induce the inclusion of a time derivative via the unsteady Stokes equations \citep{clarke2005drag}. We justify our consideration of steady Stokes flow here by noting that the oscillatory Reynolds number $\Omega L^2 / \nu$ (where $\Omega$ is the frequency of rotation, $L$ is the swimmer length, and $\nu$ is the kinematic viscosity of the fluid) is generally small for microswimmers, despite their fast self-generated motions \citep{lauga2020fluid}. Hence, while our rapid yawing analysis is relevant for the typical parameter values of many microswimmers, we emphasize that it would formally break down for very large rotation rates (e.g. $10^4$ Hz for a bacterium with lengthscale of $10 \, \mu$m in water), when induced inertial terms would become important.

The emergent dynamics for yawing in 3D cannot be understood by simply combining results for yawing in 2D and constant rotation in 3D, for two main reasons. First, 3D orientation is governed by a system of three nonlinear equations, in comparison to just a single nonlinear equation in 2D. Second, yawing corresponds to a time-dependent rather than constant angular velocity, the latter being the case for constant rotation. This means that yawing in 3D is governed by a nonautonomous nonlinear 3D dynamical system at leading order. Using a multiple scales analysis for systems, in this study we show that rapid yawing generates asymmetric emergent effects that are not present in the original particle. This emergent behaviour is fundamentally different to that arising from yawing in 2D \citep{walker2022effects} and constant rotation in 3D \citep{dalwadi2024generalisedparti,dalwadi2024generalisedpartii}. In particular, we demonstrate that the emergent asymmetry generated by rapid yawing can result in chaotic dynamics, which is not possible for passive spheroids nor for the emergent dynamics arising from rapid (constant) rotation.

We start in \S\ref{sec: Governing equations} below by setting up the physical problem and equations of motion, including in \S\ref{sec: summary of main results} a short summary of the main results we derive subsequently. We present our main analysis in \S\ref{sec: setup multiscle}, where we derive the emergent rotational dynamics of the system, relegating some of the technical details to Appendices \ref{sec: Solving LO system} and \ref{sec: Solving adjoint system}. In \S\ref{sec: chaotic behaviour}, we demonstrate that the asymmetry generated in the emergent equations can exhibit chaotic dynamics, which is a fundamentally different behaviour to that seen for passive spheroids. We then derive the emergent translational dynamics in \S\ref{sec: translational dynamics}. Finally, we discuss our results and their wider implications in \S\ref{sec: Discussion}.

\section{Problem setup}
\label{sec: Governing equations}

We consider the dynamics of a self-propelling rigid spheroid in a viscous (Stokes) fluid with an imposed far-field shear flow. We work in dimensionless quantities, scaling time with the inverse shear rate of the imposed far-field flow and space with the equatorial radius of the spheroid. The distance from the centre of the spheroid to its pole along the symmetry axis is $\ratio$, and the spheroid self-generates a fast periodic yawing within a swimmer-fixed plane containing its symmetry axis. This yawing manifests through an unsteady angular velocity $\angvel(\tstandard)$ in a quiescent fluid, where $\tstandard$ denotes time. The fast yawing means that the orientation of the swimmer varies rapidly. The spheroid also self-generates a translation $\Vel(\tstandard)$, periodic in a swimmer-fixed reference frame we define below, and with the same period as the yawing.

\begin{figure}
    \centering
    \includegraphics[width=\textwidth]{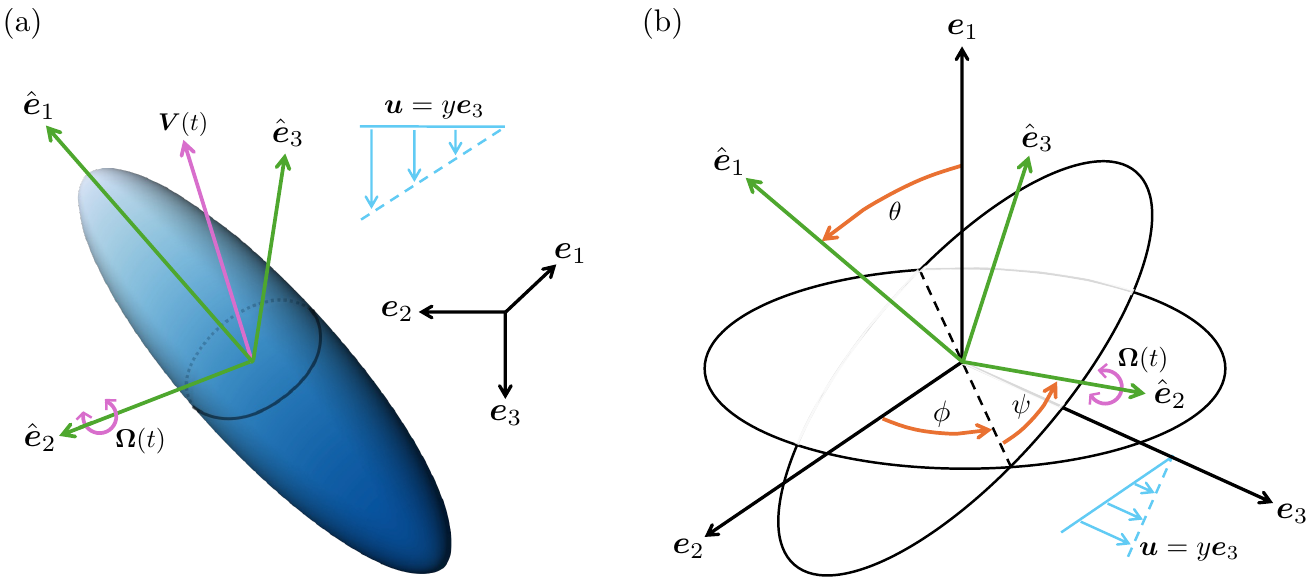}
    \caption{Schematic of (a) physical setup and (b) Euler angle definitions, with laboratory ($\e{i}$) \& swimmer-fixed ($\ehat{i}$) frames denoted by black and green arrows, respectively. The Euler angle rotations in (b) occur in the order $\phi$, $\theta$, $\psi$. The swimmer self-generates a rapid yawing via the time-dependent angular velocity $\angvel(\tstandard) = \Om \Amp \cos \left(\Om \tstandard\right) \ehat{2}$ (curved purple arrows), a time-dependent translational velocity $\Vel(\tstandard)$ (straight purple arrow in (a)), and interacts with a far-field shear flow $\flowvel = y \e{3}$ (blue arrows).}
    \label{fig: setup}
\end{figure}

We define the spheroidal axis of symmetry via a swimmer-fixed axis of $\ehat{1}$, and define the self-generated yawing through its angular velocity $\angvel(\tstandard)$, which is perpendicular to $\ehat{1}$. We then define $\ehat{2}$ to be the direction of $\angvel$, and write
\begin{align}
\label{eq: yawing motion}
\angvel(\tstandard) = \Om \Amp \cos \left(\Om \tstandard\right) \ehat{2},
\end{align}
where $\Om \gg 1$ is the fast frequency of yawing and $\Amp$ is the amplitude of yawing. Finally, we define $\ehat{3} = \ehat{1} \times \ehat{2}$. We define the orthonormal basis of the laboratory frame to be $\{\e{1},\e{2},\e{3}\}$, orientated in terms of the far-field shear flow which has velocity field $\flowvel(x,y,z) = y \e{3}$ for coordinates $(x,y,z)$ in the laboratory frame. These definitions are illustrated in Figure \ref{fig: setup}. In this swimmer-fixed basis, we also prescribe the translational velocity
\begin{align}
\label{eq: Vel def}
\Vel(\tstandard) = \sum_{i = 1}^{3} \velc_i(\tstandard) \ehat{i}, \quad \text{with } \velc_i(\tstandard) = \velav_i + \velosc_i \cos \left(\Om \tstandard - \velps_i \right),
\end{align}
where, in the $\ehat{i}$ direction, $a_i$ is the average translational velocity, $b_i$ is the amplitude of the translational velocity oscillation, and $\delta_i$ is the phase shift of this oscillation.

A key goal of our subsequent analysis is to understand the dynamics of the particle as it interacts with the far-field flow. To quantify these, we use Euler angles $\theta \in [0, \pi)$ (pitch), $\psi \text{ mod } 2\pi$ (roll), and $\phi \text{ mod } 2\pi$ (yaw), illustrated in Figure \ref{fig: setup}b, formally defined via an $xyx$-Euler angle transformation:
\begin{subequations}
\label{eq: Euler transform} 
\begin{align}
    \hat{\vec{e}}(\theta, \psi, \phi)
    = \mathsfbi{M}(\theta,\psi,\phi) \vec{e},
    \end{align}
    where $\hat{\vec{e}} := (\ehat{1}, \ehat{2}, \ehat{3})^{\intercal}$ is the swimmer-fixed basis, which depends on the orientation, $\vec{e} := (\e{1}, \e{2}, \e{3})^{\intercal}$ is the laboratory basis, and we define
    \begin{align}
     \mathsfbi{M}(\theta,\psi,\phi) := \left(\begin{array}{c|c|c}
    c_\theta & s_\phi s_\theta & - c_\phi s_\theta \\ \label{eq: M def} 
    s_\psi s_\theta &  \hphantom{+}c_\phi c_\psi  - s_\phi c_\theta s_\psi  &  \hphantom{+}s_\phi c_\psi  + c_\phi c_\theta s_\psi \\
   c_\psi s_\theta &  - c_\phi s_\psi  -s_\phi c_\theta c_\psi  & 
   -s_\phi s_\psi  + c_\phi c_\theta c_\psi
   \end{array}\right),
    \end{align}
    \end{subequations}
using the shorthand notation $s_{\theta} = \sin \theta$ etc. Then, applying the model derivation of \citet{dalwadi2024generalisedparti,dalwadi2024generalisedpartii} to the motion \eqref{eq: yawing motion}, the resulting rotational dynamics for a spheroid in shear flow with self-induced yawing are
\begin{subequations}
\label{eq: full gov eq}
\begin{align}
\label{eq: theta eq}
\dbyd{\theta}{\tstandard} &= \Om \Amp \cos \left(\Om \tstandard\right) \cos \psi + \fb(\theta,\phi; \Breth), \\
\label{eq: psi eq}
\dbyd{\psi}{\tstandard} &= - \Om \Amp \cos \left(\Om \tstandard\right) \dfrac{\cos \theta \sin \psi}{\sin \theta} + \fc(\theta,\phi; \Breth), \\
\label{eq: phi eq}
\dbyd{\phi}{\tstandard} &= \Om \Amp \cos \left(\Om \tstandard\right) \dfrac{\sin \psi}{\sin \theta} + \fa(\phi; \Breth),
\end{align}
\end{subequations}
with arbitrary initial conditions. Here, the first terms on the right-hand sides represent the fast yawing motion, and the remaining functions $\fgen$ ($i = 1, 2, 3$, here and henceforth) represent the slow interaction with the far-field shear flow. The functions $\fgen$ that quantify this interaction are
\begin{align}
\label{eq: f functions}
\fb = -\dfrac{\Breth}{2} \cos \theta \sin \theta \sin 2 \phi, \quad
\fc = \dfrac{\Breth}{2} \cos \theta \cos 2 \phi, \quad
\fa = \dfrac{1}{2} \left(1 - \Breth \cos 2 \phi\right),
\end{align}
where $\Breth(\ratio) = (\ratio^2-1)/(\ratio^2 + 1)$ is the Bretherton parameter \citep{bretherton1962motion}, which takes values $\Breth \in (-1, 1)$ for a spheroid. If there were no yawing ($\Amp = 0$), the system \eqref{eq: full gov eq} would reduce exactly to Jeffery's equations for the orientation of a passive spheroid in shear flow \citep{jeffery1922motion}. We emphasize that the right-hand sides of Jeffery's equations \eqref{eq: f functions} are independent of the spheroid roll $\psi$, illustrating their axisymmetry.

The translational dynamics for the centre of mass of the spheroid ($\bs{X} = X \e{1} + Y \e{2} + Z \e{3}$) in shear flow with self-induced translation are
\begin{align}
\label{eq: translational dynamics}
\dbyd{\bs{X}}{\tstandard} = \Vel(\tstandard) + Y \e{3},
\end{align}
with initial conditions $\bs{X}(0) = \bs{0}$, noting that we are free to prescribe the origin of the laboratory frame to be at the initial spheroid centre of mass without loss of generality. Note that $\Vel(\tstandard)$, defined in \eqref{eq: Vel def}, is only a straightforward oscillation in the swimmer frame, and that \eqref{eq: translational dynamics} is strongly coupled to the angular dynamics via the evolution of the spheroid orientation through \eqref{eq: full gov eq}. However, the reverse is not true - the angular dynamics \eqref{eq: full gov eq}--\eqref{eq: f functions} do not depend on the translational dynamics \eqref{eq: translational dynamics}. As such, it will be helpful to first consider the angular dynamics in our analysis, then to investigate the translational dynamics.

\begin{figure}
    \centering
    \includegraphics[width=\textwidth]{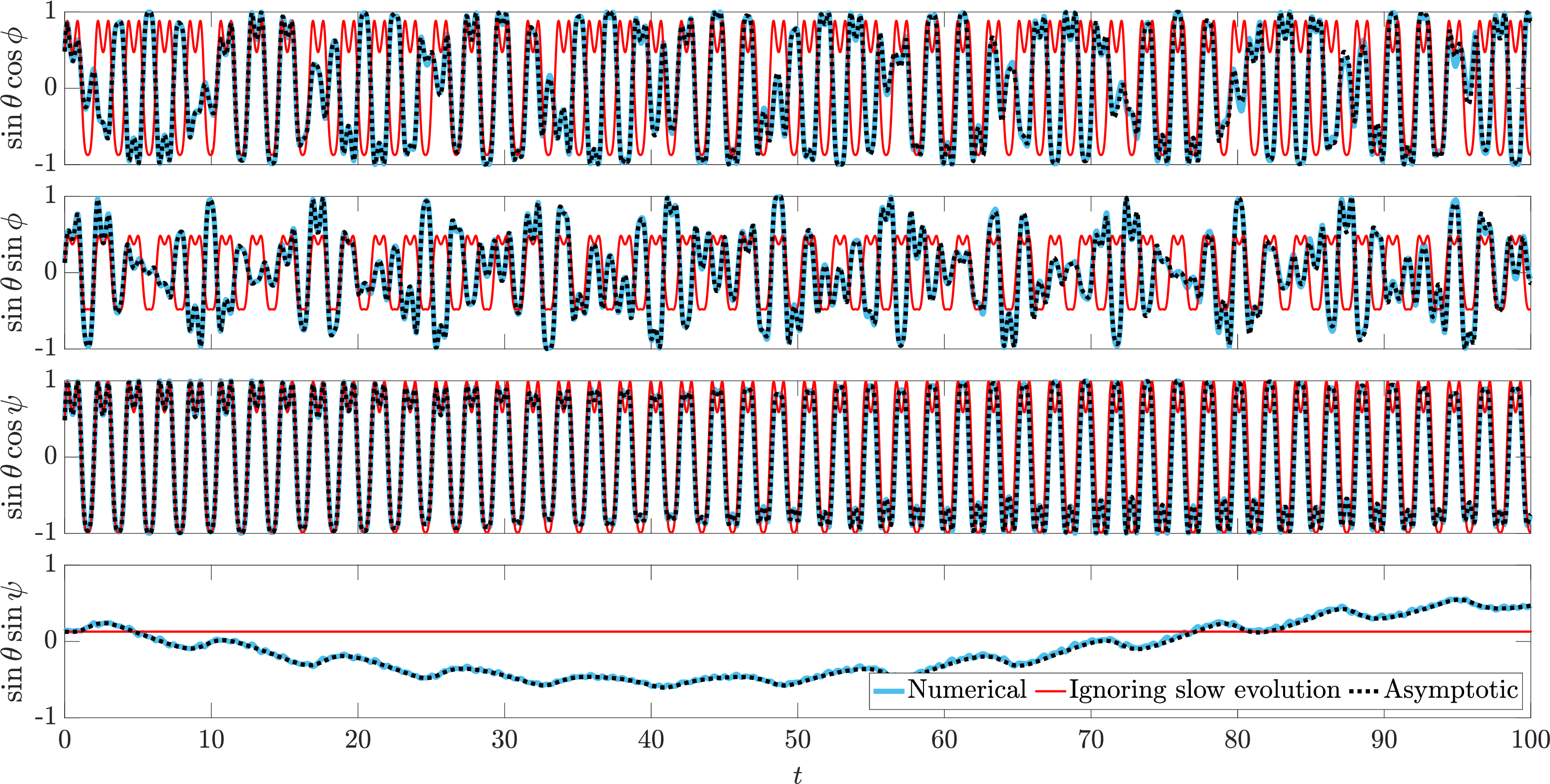}
    \caption{Numerical solutions of the full rotational dynamics \eqref{eq: full gov eq}--\eqref{eq: f functions} (solid blue lines), compared to: (1) ignoring the slow evolution, by setting $\fgen = 0$ in \eqref{eq: full gov eq} (solid red lines) and (2) the asymptotic solutions, consisting of the leading-order solutions we derive in \eqref{eq: mu def mod} and the emergent slow evolution equations we derive in \eqref{eq: slow evolution}, where the latter are solved numerically (dotted black lines). We use parameter values $\Breth = 0.9$, $\Amp = 2$, and $\Om = 3$ with initial conditions $(\theta,\psi, \phi) = (\pi/6, \pi/12, \pi/12)$. We see that the emergent (asymptotic) dynamics we derive in the limit of large $\Om$ agree well with the full dynamics, even for moderate values of $\Om$.}
    \label{fig: orientation dynamics}
\end{figure}

\begin{figure}
    \centering
    \includegraphics[width=0.9\textwidth]{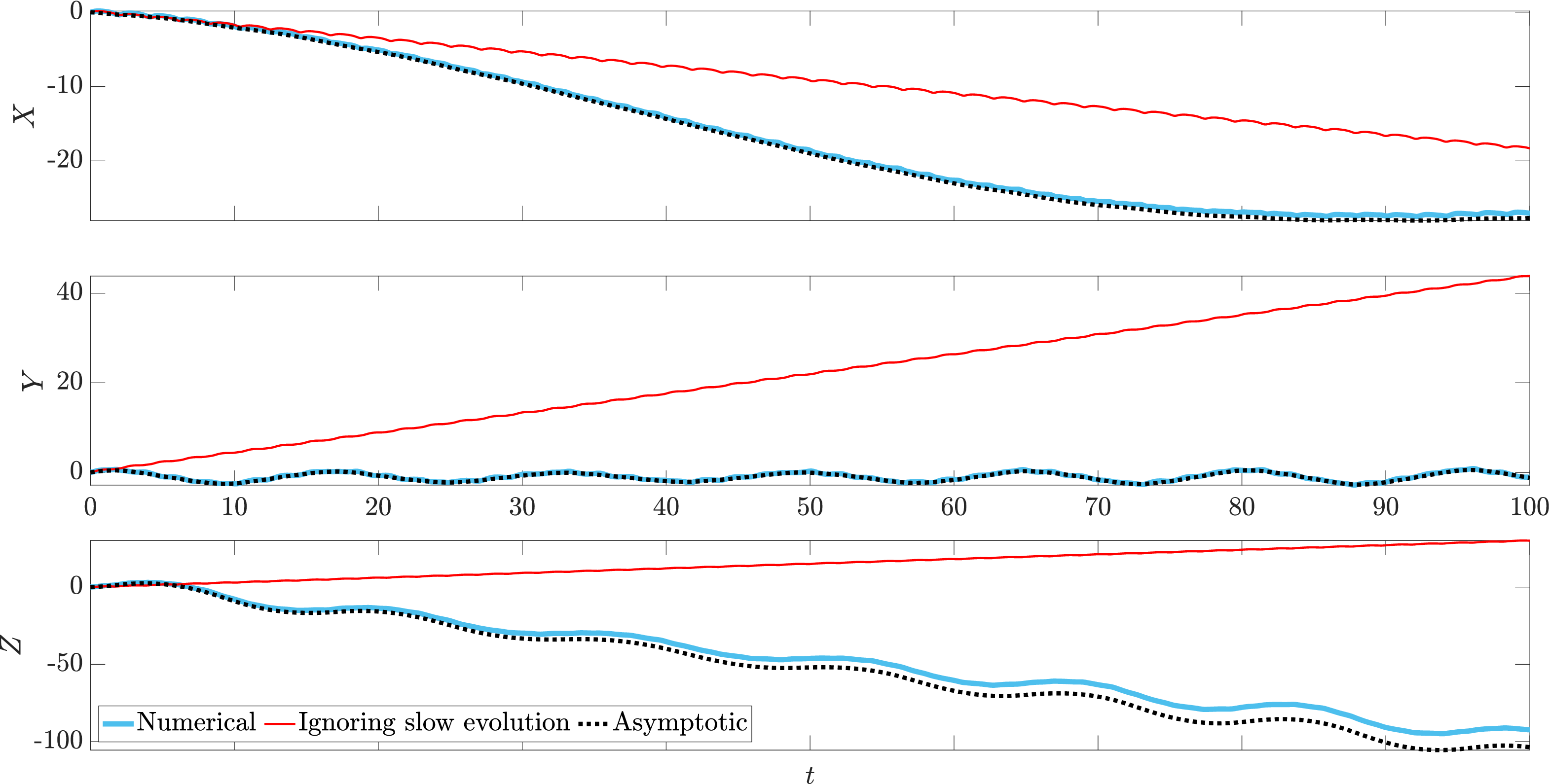}
    \caption{Numerical solutions of the full translational dynamics \eqref{eq: translational dynamics}, which also depend on the solution to the full rotational dynamics \eqref{eq: full gov eq}--\eqref{eq: f functions} (solid blue lines), compared to: (1) ignoring the slow evolution, by setting $\fgen = 0$ in \eqref{eq: full gov eq} (solid red lines) and (2) the asymptotic solutions, from the emergent slow evolution equations we derive in \eqref{eq: slow ev trans}, solved numerically (dotted black lines). We use the same parameter values as in Figure~\ref{fig: orientation dynamics}, and additionally $\Vel(\tstandard)$ is defined in \eqref{eq: Vel def}, with $(\velav_1, \velav_2, \velav_3) = (-0.2, 0.5, 0.2)$, $(\velosc_1, \velosc_2, \velosc_3) = (0.2, 0.6, 0.5)$,  $(\velps_1, \velps_2, \velps_3) = (\pi/2, \pi/4, -\pi/4)$, and initial conditions $\bs{X}(0) = \bs{0}$.}
    \label{fig: translational dynamics}
\end{figure}

The full dynamics of the nonautonomous nonlinear dynamical system \eqref{eq: full gov eq}--\eqref{eq: translational dynamics} in the fast yawing limit $\Om \gg 1$ (black lines in Figure \ref{fig: orientation dynamics}) have two main effects that occur over distinct timescales. These are: (a) yawing over a fast $\tstandard = \order{1/\Om}$ timescale, and (b) shear interaction over a slow $\tstandard = \order{1}$ timescale. In this study, we investigate the emergent dynamics of \eqref{eq: full gov eq}--\eqref{eq: translational dynamics} in the fast yawing limit where $\Om \gg 1$ (and all other parameters are of $\order{1}$). This is a singular perturbation problem where the fast oscillatory effects are maintained over the slow shear timescale i.e. the emergent dynamics cannot be obtained by simply ignoring the slow evolution due to the shear interaction (see red lines in Figures \ref{fig: orientation dynamics} and \ref{fig: translational dynamics}). We therefore use the method of multiple scales to calculate the emergent effects.

\subsection{Summary of main results}
\label{sec: summary of main results}

We show that the rotational dynamics of rapidly yawing spheroids in 3D do not act as effective passive spheroids in general, in contrast to recent results for yawing in 2D \citep{walker2022effects} and constant rotation in 3D \citep{dalwadi2024generalisedparti}. While we will show that the orientation of active spheroids here can be related to an equivalent passive particle, this effective particle only has two planes of symmetry in general, thereby losing axial symmetry. This difference in the symmetries of the effective particle is particularly notable because passive particles with two planes of symmetry have been shown to demonstrate chaotic dynamics \citep{yarin1997chaotic,thorp2019motion}. However, the classic (3D) Jeffery equations for passive spheroids are integrable \citep{jeffery1922motion}, constraining the dynamics to a 2D surface in phase space and therefore ruling out chaos \citep{thorp2019motion}.

Therefore, with the benefit of hindsight, it is helpful to record here the rotational dynamical equations for the appropriate class of asymmetric passive particles that will emerge; particles with two orthogonal planes of symmetry \citep{harris1979low,thorp2019motion,bretherton1962motion,brenner1964stokes}. These can be written as modified versions of Jeffery's equations (\eqref{eq: full gov eq} with $\Amp = 0$) transformed via
\begin{subequations}
\label{eq: g functions}
\begin{align}
\label{eq: g functions transform}
\fgen(\theta,\phi;\Breth) \mapsto \fgentransf(\theta,\psi,\phi;\Bretha,\Brethb,\Brethc) = \fgen\left(\theta,\phi;\Bretha\right) + \ggen(\theta,\psi,\phi;\Bretha,\Brethb,\Brethc),
\end{align}
where the $\ggen$ encode non-axisymmetric effects via their dependence on $\psi$, and are defined as
\begin{align}
\label{eq: g functions 1}
\gb(\theta,\psi, \phi; \Bretha, \Brethb) &=  \dfrac{\Bretha + \Brethb}{2} s_{\theta} s_{\psi}\left(c_\theta s_\psi s_{2 \phi} - c_\psi c_{2 \phi} \right), \\
\gc(\theta, \psi, \phi; \Bretha, \Brethb, \Brethc) &= \dfrac{\Bretha + \Brethb + \Brethc}{2} c_{\theta} c_{\psi} \left(c_\theta s_{\psi} s_{2 \phi} - c_{\psi} c_{2 \phi} \right) + \dfrac{\Brethc}{2} s_{\psi} \left( c_\psi s_{2 \phi} + c_\theta s_{\psi} c_{2 \phi} \right), \\
\ga(\theta,\psi, \phi; \Bretha, \Brethb) &= \dfrac{\Bretha + \Brethb}{2} \left( c^2_{\psi} c_{2 \phi} -  c_{\theta} s_{\psi} c_{\psi} s_{2\phi} \right),
\end{align}
\end{subequations}
again using the shorthand notation $s_{\theta} = \sin \theta$ etc. Importantly, the transformation \eqref{eq: g functions transform} introduces three independent coefficients $\Breth_i$ in place of $\Breth$. For an ellipsoid, the governing equations are the same as \eqref{eq: g functions} \citep{hinch1979rotation}, except that $\Breth_i$ are not independent. In fact, \citet{jeffery1922motion} showed that an ellipsoid with axes $a$, $b$, $c$ has the explicit relationship
\begin{align}
\label{eq: B_i to ratios}
\Bretha = \frac{c^2 - b^2}{c^2 + b^2}, \quad
\Brethb = \frac{a^2 - c^2}{a^2 + c^2}, \quad
\Brethc = \frac{b^2 - a^2}{b^2 + a^2},
\end{align}
from which it follows \citep{bretherton1962motion} that, for an ellipsoid, $\Breth_i$ must satisfy the relationship
\begin{align}
\label{eq: B ellipsoid relationship}
\Bretha \Brethb \Brethc + \Bretha + \Brethb + \Brethc = 0.
\end{align}
For an axisymmetric ellipsoid with $a = b$ (i.e. a spheroid), $\Bretha = -\Brethb = \Breth$ and $\Brethc = 0$, causing $\ggen= 0$ in \eqref{eq: g functions} and removing all non-axisymmetric effects, as expected.

In our analysis, we will find that the average orientation $(\newtheta,\newpsi,\newphi)$ (defined appropriately below) evolves over the $\tstandard = \order{1}$ scale, with emergent dynamics governed by
\begin{subequations}
\label{eq: slow evolution outline}
\begin{align}
\label{eq: slow evolution newtheta outline}
\dbyd{\newtheta}{\tstandard} &= \fb(\newtheta,\newphi; \Beff_1) + \gb(\newtheta,\newpsi,\newphi; \Beff_1,\Beff_2), \\
\label{eq: slow evolution newpsi outline}
\dbyd{\newpsi}{\tstandard} &= \fc(\newtheta,\newphi; \Beff_1) + \gc(\newtheta,\newpsi,\newphi; \Beff_1, \Beff_2,\Beff_3), \\
\label{eq: slow evolution newphi outline}
\dbyd{\newphi}{\tstandard} &= \fa(\newphi; \Beff_1) + \ga(\newtheta,\newpsi,\newphi; \Beff_1,\Beff_2),
\end{align}
\end{subequations}
with $\fgen$ and $\ggen$ defined in \eqref{eq: f functions}, \eqref{eq: g functions}, but now where $\Beff_i$ are functions of $\Amp$ and $\Breth(\ratio)$, and do not generally satisfy \eqref{eq: B ellipsoid relationship}. That is, we will show: (1) a rapidly yawing active spheroid generates a loss of axial symmetry in its emergent dynamics (via the $\ggen$), (2) the emergent dynamics are equivalent to those of an effective passive particle with two planes of symmetry, and (3) the effective shape is not an ellipsoid in general. The main part of our analysis involves deriving \eqref{eq: slow evolution outline} from \eqref{eq: full gov eq}--\eqref{eq: f functions}, including explicit representations of the average orientation $(\newtheta,\newpsi,\newphi)$ and the functions $\Beff_i(\Amp, \Breth)$. We also show that the emergent dynamics we derive can demonstrate chaos, and therefore that these dynamics can be fundamentally different to those for any passive spheroid.

Finally, we also determine the emergent translational dynamics over the $\tstandard = \order{1}$ scale. Though we will find that the spheroid translates with an effective velocity $\Veleff$ in addition to a straightforward shear contribution, this effective velocity is not simply the average of the physical velocity $\Vel(\tstandard)$ in the swimmer-fixed frame. Importantly, we will show that phase lags in the oscillations of the translational velocity components in the yawing plane can generate non-trivial contributions to the effective translational velocity.

\section{Deriving the emergent rotational dynamics}
\label{sec: setup multiscle}

\subsection{Framework for the method of multiple scales}
We start by analysing the emergent rotational dynamics \eqref{eq: full gov eq}--\eqref{eq: f functions} in the limit of rapid yawing, corresponding to $\Om \gg 1$. We use the method of multiple scales \citep{hinch_1991} to derive equations for the emergent behaviour. We introduce the `fast' timescale $\ts = \order{1}$ via $\ts = \Om \tl$, and refer to the original timescale $\tl$ as the `slow' timescale. Using the method of multiple scales, we treat these two timescales as independent, transforming the time derivative as
\begin{align}
\label{eq: time deriv transform}
\dbyd{}{\tstandard} \mapsto \Om \pbyp{}{\ts} + \pbyp{}{\tl}.
\end{align}
Under the time derivative mapping \eqref{eq: time deriv transform}, the dynamical system for the swimmer orientation \eqref{eq: full gov eq} is transformed to
\begin{subequations}
\label{eq: full gov eq trans General}
\begin{align}
\label{eq: theta eq trans General}
\Om \pbyp{\theta}{\ts} + \pbyp{\theta}{\tl} &= \Om \Amp \cos \ts \cos \psi + \fb(\theta,\phi;\Breth), \\
\label{eq: psi eq trans General}
\Om \pbyp{\psi}{\ts} + \pbyp{\psi}{\tl}  &= - \Om \Amp \cos \ts \dfrac{\cos \theta \sin \psi}{\sin \theta} + \fc(\theta,\phi;\Breth), \\
\label{eq: phi eq trans General}
\Om  \pbyp{\phi}{\ts} + \pbyp{\phi}{\tl} &= \Om \Amp \cos \ts \dfrac{\sin \psi}{\sin \theta} + \fa(\phi;\Breth).
\end{align}
\end{subequations}
We expand each dependent variable as an asymptotic series in inverse powers of $\Om$ and as a function of both the fast and slow timescales, writing
\begin{align}
\label{eq: asy exp}
y(\ts,\tl) \sim y_0(\ts,\tl) + \dfrac{1}{\Om} y_1(\ts,\tl) \quad \text{as } \Om \to \infty, \quad \text{ for } y \in \{\theta, \psi, \phi \}.
\end{align}

\subsection{Leading-order analysis}
\label{sec: LO}
Using the asymptotic expansions \eqref{eq: asy exp} in the transformed governing equations \eqref{eq: full gov eq trans General}, we obtain the leading-order (i.e. $\order{\Om}$) system
\begin{align}
\label{eq: full gov eq trans LO General}
\pbyp{\theta_0}{\ts}  = \Amp \cos \ts \cos \psi_0, \quad
 \pbyp{\psi_0}{\ts}  = - \Amp \cos \ts \dfrac{\cos \theta_0 \sin \psi_0}{\sin \theta_0}, \quad
 \pbyp{\phi_0}{\ts} = \Amp \cos \ts \dfrac{\sin \psi_0}{\sin \theta_0}.
\end{align}
We derive an exact solution to the nonlinear, nonautonomous leading-order system \eqref{eq: full gov eq trans LO General} in Appendix \ref{sec: Solving LO system}, obtaining
\begin{subequations}
\label{eq: mu def mod}
\begin{align}
\label{eq: om cos thet def mod}
 \cos \theta_0 &= \cos \newtheta \cos f(\ts) - \sin \newtheta \cos \newpsi \sin f(\ts), \\
\label{eq: sin theta sin psi mod}
\sin \theta_0 \sin \psi_0 &= \sin \newtheta \sin \newpsi, \\
\label{eq: tan phi0 mod}
\tan (\phi_0 - \newphi) &= \dfrac{\sin \newpsi \sin f(\ts)}{\sin \newtheta \cos f(\ts) + \cos \newtheta \cos \newpsi \sin f(\ts)},
\end{align}
\end{subequations}
defining
\begin{align}
\label{eq: f def}
f(\ts;\Amp) := \Amp \sin \ts,
\end{align}
and where $\newtheta(\tl)$, $\newpsi(\tl)$, and $\newphi(\tl)$ are the three slow-time functions of integration that remain undetermined from our leading-order analysis. The goal of the next-order analysis in \S \ref{sec: FC} will be to derive the governing equations satisfied by $\newtheta$, $\newpsi$, and $\newphi$. The three degrees of freedom that arise from integrating \eqref{eq: full gov eq trans LO General} could be included as different combinations of $\newtheta$, $\newpsi$, and $\newphi$\footnote{For example, we could have used $(\newtheta,\newpsi) \mapsto (\newtheta,H)$, where $H(\tl) = \sin \newtheta \sin \newpsi$.}; we choose the specific forms in \eqref{eq: mu def mod} so that $(\newtheta, \newpsi, \newphi)$ is associated with $(\theta, \psi, \phi)$ and represents the average orientation direction of the spheroid over a single yawing oscillation. That is $\av{\ehat{i}(\theta,\psi,\phi)} \propto \ehat{i}(\newtheta,\newpsi,\newphi)$, where we use the notation $\av{\bcdot}$ to denote the average of its argument over one fast-time oscillation, defined as
\begin{align}
\label{eq: av operator}
\av{y} = \dfrac{1}{2 \pi}\int_0^{2\pi} \! y \, \mathrm{d}\ts.
\end{align}

\subsection{Next-order system}
\label{sec: FC}

Our remaining goal is to determine the governing equations satisfied by the slow-time functions $\newtheta(\tl)$, $\newpsi(\tl)$, and $\newphi(\tl)$. To do this, we must determine the solvability conditions required for the first-order correction (i.e. $\order{1}$) terms in \eqref{eq: full gov eq trans General} after posing the asymptotic expansions \eqref{eq: asy exp}. These $\order{1}$ terms are
\begin{subequations}
\label{eq: full gov eq trans oma O1 eps}
\begin{align}
\label{eq: theta eq trans oma O1 eps}
 \pbyp{\theta_1}{\ts} + \left(\Amp \cos \ts\right) \psi_1 \sin \psi_0  &= \fb(\theta_0,\phi_0) - \pbyp{\theta_0}{\tl}, \\
\label{eq: psi eq trans oma O1 eps}
 \pbyp{\psi_1}{\ts} - \left(\Amp \cos \ts\right) \theta_1 \dfrac{\sin \psi_0}{\sin^2 \theta_0} + \left(\Amp \cos \ts\right) \psi_1 \dfrac{\cos \theta_0 \cos \psi_0}{\sin \theta_0} &= \fc(\theta_0,\phi_0) - \pbyp{\psi_0}{\tl}, \\
\label{eq: phi eq trans oma O1 eps}
 \pbyp{\phi_1}{\ts} + \left(\Amp \cos \ts\right) \theta_1 \dfrac{\cos \theta_0 \sin \psi_0}{\sin^2 \theta_0} - \left(\Amp \cos \ts\right) \psi_1\dfrac{\cos \psi_0}{\sin \theta_0} &=  \fa(\theta_0,\phi_0) - \pbyp{\phi_0}{\tl}.
\end{align}
\end{subequations}

To derive the required solvability conditions, we use the method of multiple scales for systems (see, e.g., pp.~127--128 \citet{dalwadi2014flow}, or p.~22 \citet{dalwadi2018effect}). Namely, the solvability condition for the system $L \boldsymbol{X} = \bs{G}$ can be found by calculating the solution of
\begin{align}
\label{eq: adjoint system}
L^* \bsX = \vec{0},
\end{align}
where $L^*$ is the matrix differential-algebraic linear adjoint operator. The requisite solvability condition for \eqref{eq: full gov eq trans oma O1 eps} is then
\begin{align}
\label{eq: general solv cond}
\av{\bsX \bcdot \bs{G}} = 0.
\end{align}
Generally, each linearly independent solution of the homogeneous adjoint problem \eqref{eq: adjoint system} will contribute one solvability condition. For \eqref{eq: full gov eq trans oma O1 eps}, the homogeneous adjoint operator is the transpose of the matrix operator taking the adjoint of each element, and is therefore
\begin{align}
\label{eq: adjoint matrix}
L^* = 
\begin{pmatrix} 
 - \partial_{\ts} & -\Amp \cos \ts  \sin \psi_0/\sin^2 \theta_0 & \Amp \cos \ts \cos \theta_0 \sin \psi_0 / \sin^2 \theta_0 \\
\Amp \cos \ts  \sin \psi_0 & - \partial_{\ts} + \Amp \cos \ts  \cos \theta_0 \cos \psi_0/\sin \theta_0 & - \Amp \cos \ts \cos \psi_0/\sin \theta_0 \\
0 & 0 & - \partial_{\ts} \\
\end{pmatrix}.
\end{align}
Hence, this problem is not self-adjoint i.e.~$L \neq L^*$.

Even though the system \eqref{eq: adjoint system} with operator \eqref{eq: adjoint matrix} is nonautonomous, we are able to solve it exactly by deducing appropriate nonlinear transformations. We present this analysis in Appendix \ref{sec: Solving adjoint system}, where we calculate that \eqref{eq: adjoint system}, \eqref{eq: adjoint matrix} has general periodic solution
\begin{align}
\label{eq: adjoint solution}
\bsX 
&= \Ca
\begin{pmatrix}
 \cos \theta_0 \sin \psi_0   \\ \sin \theta_0 \cos \psi_0 \\ 0
\end{pmatrix}
+ \Cb
\begin{pmatrix}
 \cos \psi_0 \\ - \sin \theta_0 \cos \theta_0 \sin \psi_0 \\ 0
\end{pmatrix}
+ \Cc
\begin{pmatrix}
0 \\ \cos \theta_0 \\ 1
\end{pmatrix}
,
\end{align}
for arbitrary constants $\Ca$, $\Cb$, and $\Cc$. The solution \eqref{eq: adjoint solution} can also be verified \emph{a posteriori} by direct substitution into \eqref{eq: adjoint system}, \eqref{eq: adjoint matrix} and applying \eqref{eq: full gov eq trans LO General}.

Finally, we derive our required solvability conditions by substituting the adjoint solutions \eqref{eq: adjoint solution} into the general solvability condition \eqref{eq: general solv cond}, with $\bs{G}$ defined as the vector right-hand side of \eqref{eq: full gov eq trans oma O1 eps}, and setting the resulting coefficients of $\Ci$ to zero. This procedure yields the following three solvability conditions
\begin{subequations}
\label{eq: solv conditions orig}
\begin{align}
\label{eq: solv conditions orig fb fc 1}
\av{\theta_{0 \tl} \cos \theta_0 \sin \psi_0  + \psi_{0 \tl} \sin \theta_0 \cos \psi_0} &= \av{\fb \cos \theta_0 \sin \psi_0 +  \fc \sin \theta_0 \cos \psi_0}, \\
\label{eq: solv conditions orig fb fc 2}
\av{\theta_{0 \tl} \cos \psi_0 -\psi_{0 \tl} \sin \theta_0 \cos \theta_0 \sin \psi_0} &= \av{\fb \cos \psi_0 - \fc \sin \theta_0 \cos \theta_0 \sin \psi_0}, \\
\label{eq: solv conditions orig fa fc}
\av{\psi_{0\tl} \cos \theta_0 + \phi_{0\tl}} &= \av{\fc \cos \theta_0 + \fa},
\end{align}
\end{subequations}
where the subscript $\tl$ denotes partial differentiation with respect to $\tl$. To derive the emergent governing equations we seek, our remaining task is to evaluate the averages in \eqref{eq: solv conditions orig} in terms of the slow-term functions $\newtheta$, $\newpsi$, and $\newphi$.

\subsection{Evaluating the solvability conditions}

Using our leading-order solutions \eqref{eq: mu def mod}, we evaluate the left-hand sides of \eqref{eq: solv conditions orig} to obtain
\begin{subequations}
\label{eq: solv conditions orig mod LHS}
\begin{align}
\label{eq: solv conditions orig fb fc 1 mod LHS}
\av{\theta_{0 \tl} \cos \theta_0 \sin \psi_0  + \psi_{0 \tl} \sin \theta_0 \cos \psi_0} &= \newtheta_\tl \cos \newtheta \sin \newpsi + \newpsi_\tl \sin \newtheta \cos \newpsi, \\
\label{eq: solv conditions orig fb fc 2 mod LHS}
\av{\theta_{0 \tl} \cos \psi_0 -\psi_{0 \tl} \sin \theta_0 \cos \theta_0 \sin \psi_0} &= \newtheta_\tl \cos \newpsi - \newpsi_\tl \sin \newtheta \cos \newtheta \sin \newpsi, \\
\label{eq: solv conditions orig fa fc mod LHS}
\av{\psi_{0\tl} \cos \theta_0 + \phi_{0\tl}} &= \newpsi_\tl \cos \newtheta + \newphi_\tl.
\end{align}
\end{subequations}
After more algebra, we can also evaluate the right-hand sides of \eqref{eq: solv conditions orig} to deduce that
\begin{subequations}
\label{eq: solv conditions orig RHS}
\begin{align}
\label{eq: solv conditions orig fb fc 1 RHS}
&\fb \cos \theta_0 \sin \psi_0 +  \fc \sin \theta_0 \cos \psi_0 \notag \\
&\qquad = -\dfrac{\Breth}{2} \left\{s_{\newtheta} s_{\newpsi} \left[c_{\newtheta}^2 c_f^2 - c_{\newpsi}^2 (1 + c_{\newtheta}^2)s_f^2 \right] + c_{\newtheta}^3 s_{2 \newpsi} s_f c_f \right\} s_{2 \newphi} \notag \\
&\qquad \quad -\dfrac{\Breth}{2} \left\{ s_{\newtheta} c_{\newtheta} c_{\newpsi} \left( c_{2\newpsi} s_f^2 - c_f^2\right) + \left[s_{\newtheta}^2 c_{\newpsi}^2 - c_{\newtheta}^2 c_{2 \newpsi} \right] s_f c_f \right\} c_{2 \newphi} , \\
\label{eq: solv conditions orig fb fc 2 RHS}
&\fb \cos \psi_0 - \fc \sin \theta_0 \cos \theta_0 \sin \psi_0 \notag \\
&\qquad = \dfrac{\Breth}{2} \left\{s_{\newtheta} c_{\newtheta} c_{\newpsi} \left(c_{2 \newpsi} s_f^2 - c_f^2 \right) + \left[s_{\newtheta}^2 c_{\newpsi}^2 - c_{\newtheta}^2 c_{2 \newpsi} \right] s_f c_f \right\} s_{2 \newphi} \notag \\
&\qquad \quad-\dfrac{\Breth}{2} \left\{s_{\newtheta} s_{\newpsi} \left[ c_{\newtheta}^2 c_f^2 - c_{\newpsi}^2 (1 + c_{\newtheta}^2) s_f^2 \right] +  c_{\newtheta}^3 s_{2 \newpsi} s_f c_f \right\} c_{2 \newphi}, \\
\label{eq: solv conditions orig fa fc RHS}
&\fc \cos \theta_0 + \fa =  \dfrac{1}{2} \left(1 - \Breth \left\{ (c_{\newtheta}^2 c_{\newpsi}^2 - s_{\newpsi}^2) s_f^2 + s_{\newtheta}^2 c_f^2 + s_{\newtheta} c_{\newtheta} c_{\newpsi} c_f s_f \right\}c_{2 \newphi}\right)  \notag \\
&\qquad \qquad \qquad \, \, \,  + \Breth s_{\newpsi} [c_{\newtheta} c_{\newpsi} s_f^2 + s_{\newtheta}^2 c_f s_f] s_{2 \newphi} \} ),
\end{align}
\end{subequations}
using the shorthand notation $s_{\newtheta} = \sin \newtheta$ etc, and where $f$ is defined in \eqref{eq: f def}. Then, noting that $\av{c_f} = J_0(\Amp)$ and $\av{s_f} = 0$, deduced via the integral representation of $J_0(\Amp)$ (the Bessel function of order zero) and parity arguments, we obtain
\begin{align}
\label{eq: integrals}
\av{s_f^2} = \dfrac{1}{2}\left(1 - J_0(2 \Amp) \right), \quad
\av{c_f^2} = \dfrac{1}{2}\left(1 + J_0(2 \Amp) \right), \quad
\av{s_f c_f} = 0.
\end{align}

\begin{figure}
    \centering
    \includegraphics[width=\textwidth]{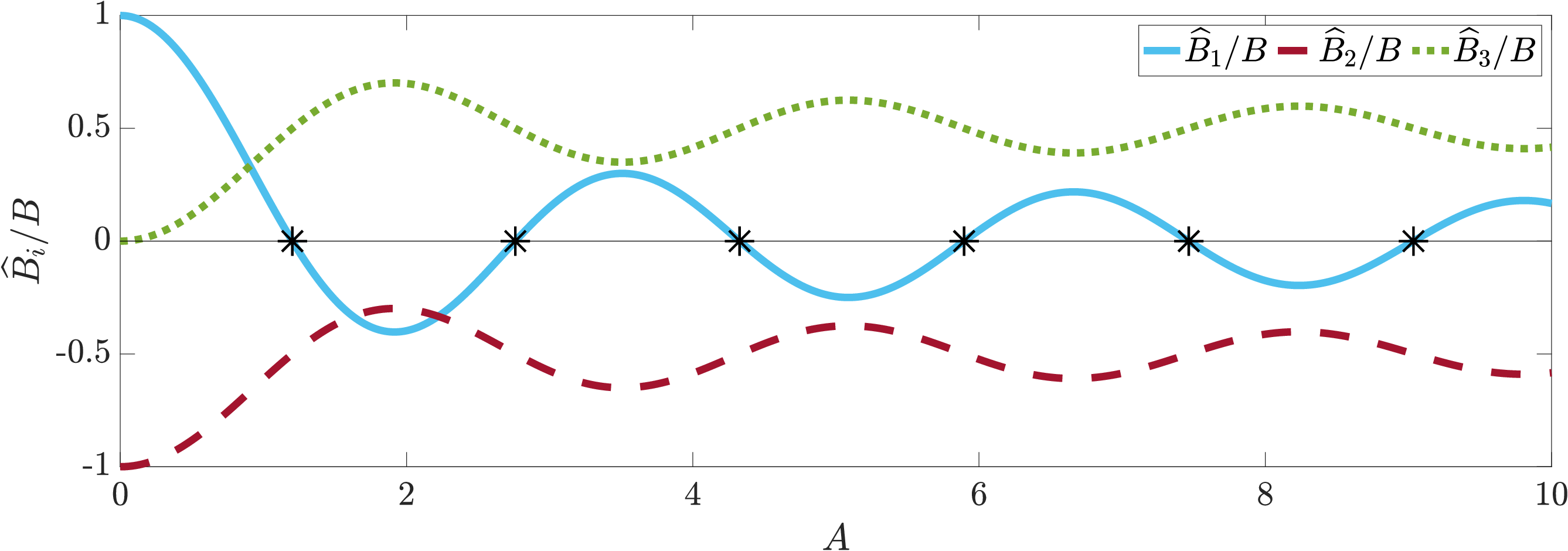}
    \caption{The effective coefficients \eqref{eq: Effective coefficients}, obtained by comparing \eqref{eq: slow evolution} to \eqref{eq: slow evolution outline}. The marked stars along the $x$-axis are the values of $\Amp$ at which \eqref{eq: B effective ellipsoid relationship} is satisfied (i.e. $J_0(2 \Amp) = 0$) and therefore when the effective shape is an ellipsoid. In fact, in these cases the effective shape is constrained further to a spheroid, whose aspect ratio is given in \eqref{eq: effective aspect ratio}.}
    \label{fig: effective coefficients}
\end{figure}

Taking the averages of \eqref{eq: solv conditions orig RHS} using \eqref{eq: integrals} allows us to evaluate the right-hand sides of the solvability conditions \eqref{eq: solv conditions orig}. Then, combining this result with \eqref{eq: solv conditions orig mod LHS} for the left-hand sides of \eqref{eq: solv conditions orig} and rearranging, we obtain
\begin{subequations}
\label{eq: slow evolution}
\begin{align}
\label{eq: slow evolution newtheta}
\dbyd{\newtheta}{\tl} &= -\dfrac{\Breth J_0(2\Amp)}{2} s_{\newtheta} c_{\newtheta} s_{2\newphi} -\dfrac{\Breth (1 - J_0(2\Amp))}{4} s_{\newtheta} s_{\newpsi} \left(c_{\newtheta} s_{\newpsi} s_{2 \newphi} - c_{\newpsi} c_{2 \newphi} \right), \\
\label{eq: slow evolution newpsi}
\dbyd{\newpsi}{\tl} &= \dfrac{\Breth J_0(2\Amp)}{2} c_{\newtheta} c_{2 \newphi} + \dfrac{\Breth (1 - J_0(2 \Amp))}{4} s_{\newpsi} \left( c_{\newpsi} s_{2 \newphi} + c_{\newtheta} s_{\newpsi} c_{2 \newphi} \right), \\
\label{eq: slow evolution newphi}
\dbyd{\newphi}{\tl} &= \dfrac{1}{2} \left(1 - \Breth J_0(2\Amp) c_{2 \newphi}\right) - \dfrac{\Breth (1 - J_0(2\Amp))}{4} c_{\newpsi} \left(c_{\newpsi} c_{2 \newphi} - c_{\newtheta} s_{\newpsi} s_{2 \newphi} \right),
\end{align}
\end{subequations}
which are the key results of our analysis; the emergent slow evolution equations we have been seeking. Recombining the slow evolution equations \eqref{eq: slow evolution} with the fast oscillations \eqref{eq: mu def mod}, we see the leading-order solutions agree very well with the full dynamics (blue \& black lines in Figure \ref{fig: orientation dynamics}), even for values of $\Om$ as low as $3$. The agreement improves further as $\Om$ increases. Finally, we note that the emergent equations \eqref{eq: slow evolution} have the functional form we claimed in \eqref{eq: slow evolution outline}, with analytically derived representations for the coefficients (illustrated in Figure \ref{fig: effective coefficients}):
\begin{align}
\label{eq: Effective coefficients}
\Beff_1 = \Breth J_0(2 \Amp), \quad
\Beff_2 = -\dfrac{\Breth}{2} \left(1 + J_0(2 \Amp) \right), \quad
\Beff_3 = \dfrac{\Breth}{2} \left(1 - J_0(2 \Amp) \right).
\end{align}

To summarise, we have demonstrated that the emergent rotational dynamics \eqref{eq: slow evolution} gain an effective asymmetry, with effective coefficients derived in \eqref{eq: Effective coefficients}. This is most explicitly demonstrated through their dependence on $\newpsi$ in the terms proportional to $\Breth (1 - J_0(2\Amp))$ on the right-hand sides of \eqref{eq: slow evolution}. As discussed in \S \ref{sec: summary of main results}, the emergent dynamics \eqref{eq: slow evolution} are equivalent to those of a passive object with two orthogonal planes of symmetry. If the effective coefficients \eqref{eq: Effective coefficients} were independent, then it would immediately follow from the results of \citet{yarin1997chaotic,thorp2019motion} that chaotic behaviour was possible. Since the effective coefficients \eqref{eq: Effective coefficients} are not independent, it is not immediately clear whether such behaviour is also possible in the system we derive. In the next section, we demonstrate that chaotic behaviour is possible in the system \eqref{eq: slow evolution}--\eqref{eq: Effective coefficients}.

\section{Chaotic behaviour in the emergent dynamics}
\label{sec: chaotic behaviour}

To investigate the possibility of chaos in the system \eqref{eq: slow evolution}--\eqref{eq: Effective coefficients}, we follow \citet{hinch1979rotation,yarin1997chaotic,thorp2019motion} and define an appropriate Poincar\'e section. Specifically, we reduce the full 3D continuous dynamics of \eqref{eq: slow evolution}--\eqref{eq: Effective coefficients} to a 2D discrete dynamical system in $(\newpsi,\newtheta)$ whenever $\newphi = n \pi$ for non-negative integer $n$. We can do this straightforwardly by solving \eqref{eq: slow evolution newtheta}--\eqref{eq: slow evolution newpsi} in terms of $\newphi$, transforming the time derivatives via
\begin{align}
\dbyd{}{\tl} \mapsto \dbyd{\newphi}{\tl} \dbyd{}{\newphi},
\end{align}
and using \eqref{eq: slow evolution newphi} to evaluate $\mathrm{d} \newphi/ \mathrm{d} \tl >0$. The monotonic nature of $\mathrm{d} \newphi/ \mathrm{d} \tl$ follows from $|\Breth| < 1$, and ensures that the transformation is well defined. For given system parameters $\Amp$ (amplitude of yawing), $\Breth = (\ratio^2 - 1)/(\ratio^2 + 1)$ (Bretherton parameter in terms of spheroid aspect ratio, $\ratio$), and initial conditions $(\newpsi_0, \newtheta_0) = (\newpsi(0), \newtheta(0))$ (setting $\newphi(0) = 0$), this generates an iteration $(\newpsi_n,\newtheta_n)$.

\begin{figure}
    \centering
    \includegraphics[width=0.7\textwidth]{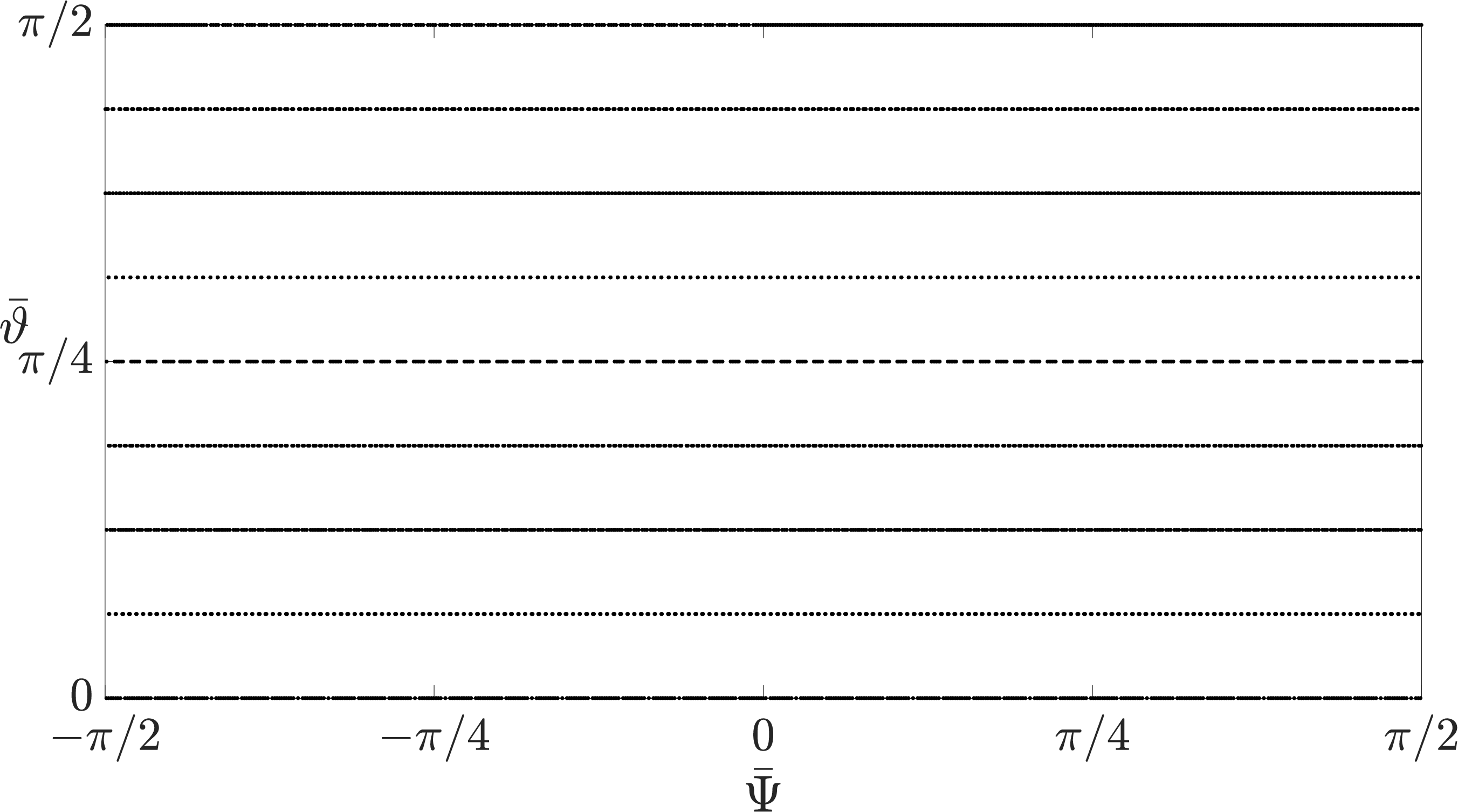}
    \caption{Poincar\'e section for the classic Jeffery's equations, which are equivalent to setting $\Amp = 0$ in our emergent equations \eqref{eq: slow evolution}--\eqref{eq: Effective coefficients}. Using the Poincar\'e map outlined in the main text, we use $\Amp = 0$, $\Breth = 0.99$ and iterate up to $n = 500$. The full 2D phase space is obtained by exploiting reflectional symmetry across $\newtheta = \pi/2$ and translational symmetry in $\newpsi \mapsto \newpsi + \pi$. No chaos is possible for the classic Jeffery's equations, as observed here.}
    \label{fig: Poincare classic Jeffery}
\end{figure}

\begin{figure}
    \centering
    \includegraphics[width=\textwidth]{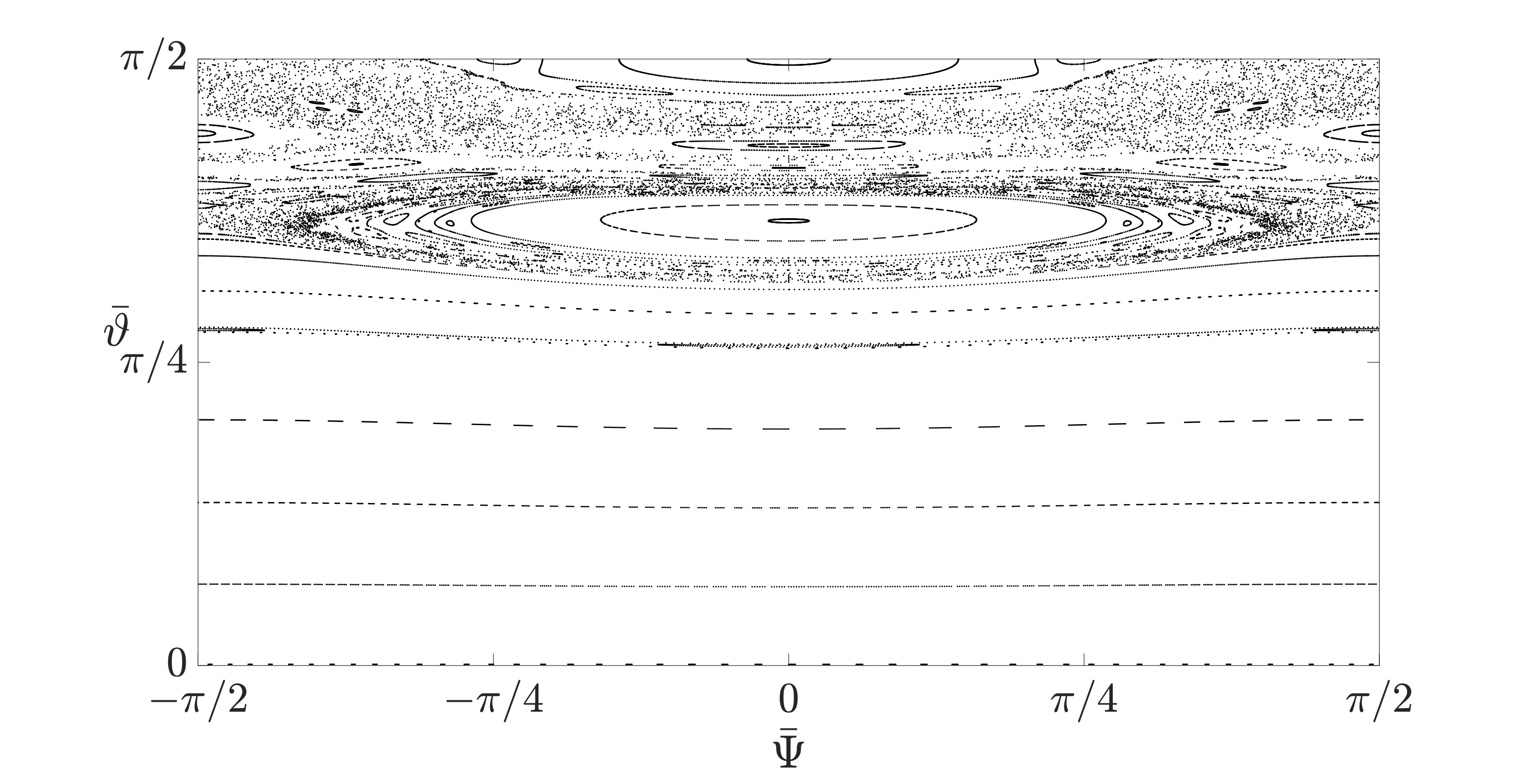}
    \caption{Poincar\'e section for the emergent dynamics \eqref{eq: slow evolution}--\eqref{eq: Effective coefficients}. Using the Poincar\'e map outlined in the main text, we use $\Amp = 0.25$, $\Breth = 0.99$ and iterate up to $n = 500$. The full 2D phase space is obtained by exploiting reflectional symmetry across $\newtheta = \pi/2$ and translational symmetry in $\newpsi \mapsto \newpsi + \pi$. Significant regions of chaos are observed in the upper third of the figure.}
    \label{fig: Poincare detailed}
\end{figure}

In Figures \ref{fig: Poincare classic Jeffery} and \ref{fig: Poincare detailed}, we show Poincar\'e sections for $\Amp = 0$ (i.e. classic Jeffery orbits) and $A = 0.25$, respectively. These consist of a point shown for every iteration up to $n=500$ for different initial conditions. Periodic orbits in $(\newpsi_n,\newtheta_n)$ repeat exactly, and are represented by distinct points that repeat themselves after a finite number of iterations. Quasiperiodic orbits appear as 1D curves i.e. orbits consist of points that densely cover a 1D path but never exactly repeat themselves. These correspond to orbits with more than one periodic component, but with incommensurate frequencies. Finally, chaotic orbits appear as dense patchworks of points in a 2D region of the Poincar\'e section.

For the classic Jeffery's equations (for passive spheroids), the orbits are constant in $\newtheta$ (Figure \ref{fig: Poincare classic Jeffery}). No chaos is possible in this case, which follows from the integrability of the classic 3D Jeffery's equations constraining the dynamics to a 2D surface in phase space. The lack of chaotic dynamics for passive spheroids can be seen visually in Figure \ref{fig: Poincare classic Jeffery}. These features are well known for passive spheroids, and have been explored in detail as part of the more general analysis in \citet{thorp2019motion}.

In contrast to the classic Jeffery's equations, we see that chaos is possible for the more general emergent equations we derive here (Figure \ref{fig: Poincare detailed}). Chaotic regions are plentiful in the upper third of the figure. In the lower two-thirds of the figure, the behaviour is mainly dominated by periodic and quasiperiodic orbits, and we note that these behaviours also appear within islands in the upper third. The quasiperiodic behaviours include orbits that sample all values of $\newpsi \in [- \pi/2,  \pi/2]$ and orbits that only take a subset of values therein.

\begin{figure}
    \centering
    \includegraphics[width=\textwidth]{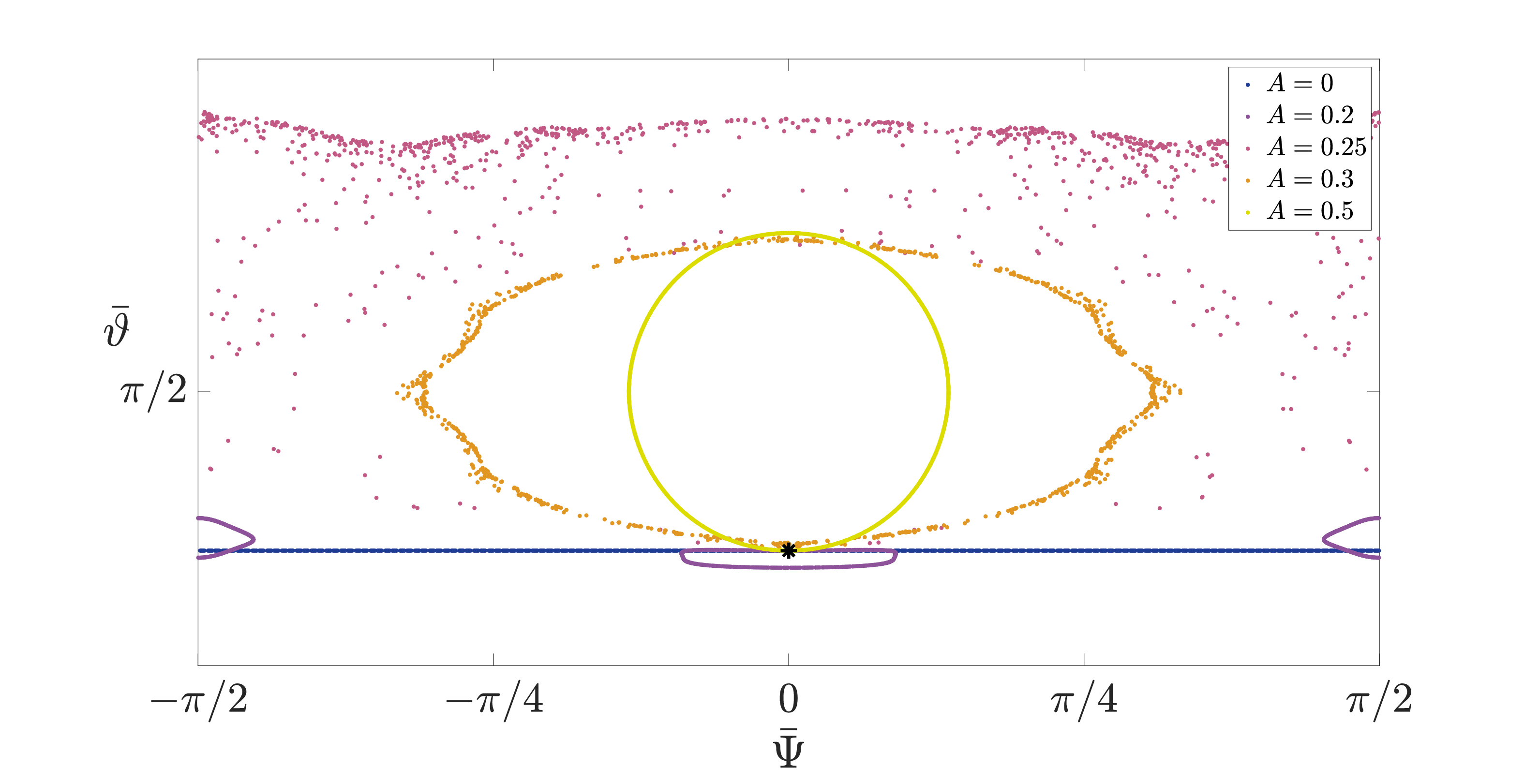}
    \caption{Poincar\'e section for the emergent dynamics \eqref{eq: slow evolution}--\eqref{eq: Effective coefficients}. Using the Poincar\'e map outlined in the main text, we use $\Breth = 0.99$ with initial conditions $(\newpsi_0, \newtheta_0) = (0, 9\pi/20)$ (shown as a black asterisk), and iterate up to $n = 1000$. We use $\Amp \in \{0, 0.2, 0.25, 0.3, 0.5 \}$, as described in the legend. For lower values of $\Amp$ the orbits are quasiperiodic. However, the orbit is chaotic at $\Amp = 0.25$ and $\Amp = 0.3$, before returning to quasiperiodicity for $\Amp = 0.5$.}
    \label{fig: Poincare changing A}
\end{figure}

Finally, in Figure \ref{fig: Poincare changing A}, we also show Poincar\'e sections for different values of $\Amp$, starting with the same initial condition. For lower values of $\Amp$, the orbits are quasiperiodic. However, the orbits demonstrate chaos as $\Amp$ increases, before returning to a quasiperiodic orbit as $\Amp$ increases further.

Hence, we have shown that chaotic behaviour is possible in the emergent dynamics we derive \eqref{eq: slow evolution}--\eqref{eq: Effective coefficients}. This behaviour arises as a direct result of the asymmetry that is generated by the rapid yawing of the active spheroids we consider. This asymmetry does not arise from a rapid (constant) rotation of spheroids, for which the emergent dynamics are equivalent to equivalent passive spheroids. In this latter case, integrability of the system means that chaotic behaviour is not possible, as shown explicitly in \citet{thorp2019motion}.

\section{Deriving the emergent translational dynamics}
\label{sec: translational dynamics}

We now consider the emergent translational dynamics from the system \eqref{eq: translational dynamics} in the large-$\Om$ limit. The emergent translational dynamics we will derive are significantly more straightforward than their rotational equivalents. In multiple scales form, the time derivative is again transformed via \eqref{eq: time deriv transform}, and so \eqref{eq: translational dynamics} becomes
\begin{align}
\label{eq: MoMS translation}
\Om \pbyp{\bs{X}}{\ts} + \pbyp{\bs{X}}{\tl} = \Vel(\ts) + Y \e{3},
\end{align}
emphasizing that $\bs{X} = \bs{X}(\ts,\tl)$ in general, and explicitly writing the velocity $\Vel$ from \eqref{eq: Vel def} in terms of the fast timescale as
\begin{align}
\label{eq: Vel def fast}
\Vel(\ts) = \sum_{i = 1}^{3} \velc_i(\ts) \ehat{i}, \quad \text{where } \velc_i(\ts) := \velav_i + \velosc_i \cos \left(\ts - \velps_i \right).
\end{align}
Expanding $\bs{X} \sim \bs{X}_0 + (1/\Om) \bs{X}_1$ and substituting into \eqref{eq: MoMS translation}, the $\order{\Om}$ terms yield
\begin{align}
\pbyp{\bs{X}_0}{\ts} = \bs{0},
\end{align}
which is trivially solved by $\bs{X}_0 = \bs{X}_0(\tl)$. At next order, the $\order{1}$ terms in \eqref{eq: MoMS translation} yield
\begin{align}
\label{eq: Next order translational dynamics}
\pbyp{\bs{X}_1}{\ts} + \dbyd{\bs{X}_0}{\tl} = \Vel(\ts) + Y_0 \e{3},
\end{align}
where we note that the dependence of $\Vel$ on $\ehat{i}$ in \eqref{eq: Vel def fast} means that $\Vel$ depends on the leading-order Euler angles. These are defined via the fast oscillations \eqref{eq: mu def mod} and the slow evolution equations \eqref{eq: slow evolution}.

The emergent equations are obtained by averaging \eqref{eq: Next order translational dynamics} over a $2 \pi$ period in $\ts$. Performing this procedure, the fast time derivative of $\bs{X}_1$ vanishes (by design), and we are left with
\begin{align}
\label{eq: slow ev trans midway}
\dbyd{\bs{X}_0}{\tl} = \av{\sum_{i = 1}^{3} \left[\velav_i + \velosc_i \cos \left(\ts - \velps_i \right) \right] \ehat{i}(\theta_0, \psi_0, \phi_0)} + Y_0 \e{3},
\end{align}
where we have replaced $\Vel(\ts)$ in the average operator with its specific form via \eqref{eq: Vel def fast}, to emphasize the dependence of $\Vel(\ts)$ on the leading-order spheroid orientation. To evaluate this averaged velocity, we use the transformation between laboratory basis and swimmer-fixed basis \eqref{eq: Euler transform} to write $\ehat{i}$ explicitly in terms of $(\theta_0, \psi_0, \phi_0)$ and the laboratory basis. Then, we combine this with our leading-order fast-time solutions \eqref{eq: mu def mod} and directly calculate the following averages:
\begin{subequations}
\label{eq: av translation results}
\begin{align}
\label{eq: av swimmer frame}
\av{\hat{\vec{e}}(\theta_0, \psi_0, \phi_0)} &= \mathsfbi{D}\, \tilde{\vec{e}}(\newtheta,\newpsi,\newphi), \\
\av{\hat{\vec{e}}(\theta_0, \psi_0, \phi_0) \sin \ts} &= \mathsfbi{A} \, \tilde{\vec{e}}(\newtheta,\newpsi,\newphi), \\
\av{\hat{\vec{e}}(\theta_0, \psi_0, \phi_0) \cos \ts} &= \bs{0},
\end{align}
where
\begin{align}
\mathsfbi{D} = \left(\begin{array}{c c c}
    J_0(\Amp) & 0 & 0 \\
    0 & 1 & 0 \\
   0 & 0  & J_0(\Amp)
   \end{array}\right), \quad
   \mathsfbi{A} = \left(\begin{array}{c c c}
    0 & 0 & J_1(\Amp) \\
    0 & 0 & 0 \\
   -J_1(\Amp) & 0  & 0
   \end{array}\right), \quad
   \tilde{\vec{e}}(\newtheta,\newpsi,\newphi) = \mathsfbi{M}(\newtheta,\newpsi,\newphi) \vec{e},
\end{align}
\end{subequations}
noting the appearance of Bessel functions of order one in $\mathsfbi{A}$, along with the Bessel functions of order zero in $\mathsfbi{D}$ that we have already seen appear in the rotational dynamics (though with a slightly different argument). Before we use these results to determine the average translational dynamics, it is instructive to note that $\tilde{\vec{e}} = (\tilde{\vec{e}}_1, \tilde{\vec{e}}_2, \tilde{\vec{e}}_3)^{\intercal}$ is defined as the basis transformation $\mathsfbi{M}$ (defined in \eqref{eq: Euler transform}) applied to the laboratory basis, but evaluated using the slow evolution angles $(\newtheta,\newpsi,\newphi)$ instead of the rapidly varying full Euler angles $(\theta, \psi, \phi)$. That is, \eqref{eq: av swimmer frame} tells us that the average of the leading-order swimmer-fixed basis $\av{\hat{\vec{e}}}$ is the slow evolution angle basis, weighted by the diagonal matrix $\mathsfbi{D}$. We can therefore, in some sense, interpret $(\newtheta,\newpsi,\newphi)$ as an `average orientation' of the spheroid over the fast timescale.

Using the averaged results \eqref{eq: av translation results} in the solvability condition \eqref{eq: slow ev trans midway}, we obtain the emergent governing equation for the translational dynamics
\begin{subequations}
\label{eq: slow ev trans}
\begin{align}
\dbyd{\bs{X}_0}{\tl} = \Veleff + Y_0 \e{3},
\end{align}
where the effective self-generated translational velocity $\Veleff$ is defined as
\begin{align}
\label{eq: Veleff}
\Veleff = \left[\velav_1 J_0(\Amp) + \velosc_3 J_1(\Amp) \sin \velps_3 \right] \tilde{\vec{e}}_1
+ \velav_2  \tilde{\vec{e}}_2
+ \left[\velav_3 J_0(\Amp) - \velosc_1 J_1(\Amp) \sin \velps_1 \right] \tilde{\vec{e}}_3.
\end{align}
\end{subequations}
Importantly, we see that the effective velocity is constant in the averaged swimmer-fixed frame defined through $\tilde{\vec{e}}(\newtheta,\newpsi,\newphi)$. We discuss the implication of these results in \S \ref{sec: Discussion}.

\section{Discussion}
\label{sec: Discussion}

We show that a rapidly yawing spheroidal swimmer interacting with a far-field shear flow generates non-axisymmetric emergent effects in its rotational dynamics, equivalent to those of a passive particle with two orthogonal planes of symmetry. With the caveat that the effective coefficients we derive \eqref{eq: Effective coefficients} are not independent, our emergent equations are equivalent to those that have been investigated in e.g. \citet{hinch1979rotation,thorp2019motion,yarin1997chaotic}. From these works, it is known that passive particles with this type of asymmetry can behave very differently to passive spheroids. We demonstrate that the emergent dynamics we derive here can exhibit chaotic behaviours, in stark contrast to passive spheroids.

Given that the effective coefficients we derive \eqref{eq: Effective coefficients} are not independent (notably, $\Beff_1 + \Beff_2 + \Beff_3 = 0$), the effective shape described by our emergent equations is constrained within the space of particles with two orthogonal planes of symmetry. A basic class of objects with this type of symmetry is ellipsoids. To check when our effective coefficients describe a passive ellipsoid, we use \eqref{eq: B ellipsoid relationship} to derive the requirement
\begin{align}
\label{eq: B effective ellipsoid relationship}
0 = \Beff_1 \Beff_2 \Beff_3 + \Beff_1 + \Beff_2 + \Beff_3 = -\Breth^3 J_0(2 \Amp) \left(1 - J_0^2(2 \Amp) \right)/4.
\end{align}
Therefore, the equivalent effective particle described by the emergent evolution equations \eqref{eq: slow evolution} is not an ellipsoid in general, unless one of three specific conditions hold: (1) $\Amp = 0$ (i.e. no yawing), (2) $\Breth = 0$ (i.e. $\ratio = 1$; the original spheroid is a sphere), or (3) $J_0(2 \Amp) = 0$. There are infinitely many discrete values of $\Amp$ that generate the non-trivial case (3); for these scenarios we can use the relationship \eqref{eq: B_i to ratios} between $\Breth_i$ and the ellipsoid axes to deduce that the effective passive shape becomes a spheroid with axes $\widehat{a}$, $\widehat{b}$, $\widehat{b}$, where
\begin{align}
\label{eq: effective aspect ratio}
\widehat{a}/\widehat{b} = \sqrt{(\ratio^2 + 3)/(3 \ratio^2 + 1)}.
\end{align}
Thus, in case (3) active prolate spheroids behave as passive oblate spheroids and vice versa. Notably, the effective aspect ratio \eqref{eq: effective aspect ratio} is the same as that which arises for a spheroid rapidly and uniformly rotating about an axis perpendicular to its symmetry axis \citep{dalwadi2024generalisedparti}. This can be understood intuitively in the large-$\Amp$ limit, where $J_0(2 \Amp) \to 0$, for which the large-amplitude yawing has a similar effect to uniform rotation. However, we emphasize that case (3) also occurs for the infinitely many finite roots of $J_0(2 \Amp) = 0$.

The emergent loss of symmetry here is fundamentally different to the results of recent studies of different types of rapidly moving rigid bodies e.g. yawing in 2D \citep{walker2022effects}, and constant rotation in 3D \citep{dalwadi2024generalisedparti,dalwadi2024generalisedpartii}. While these studies do also show that their specific rapid motions in shear flow lead to emergent dynamics, the effective passive shapes generated all preserve the hydrodynamic symmetries of the original physical shapes. Moreover, the equivalence to an effective passive spheroid is generic for periodically shape-deforming swimmers in 2D \citep{gaffney2022canonical}. Our study shows that symmetry of the physical swimmer is not maintained for rapid yawing in 3D.

Given the fundamental difference between our 3D results \eqref{eq: slow evolution} and the generic 2D behaviour \citep{gaffney2022canonical}, it is instructive to understand how our results collapse to the 2D case. By constraining the swimmer pitch and yawing motions to the shear plane ($\theta = \psi = \pi/2$ in the full dynamics \eqref{eq: full gov eq}) and solely evolving the remaining equation for $\phi$, we reduce our setup to the 2D yawing problem considered in \citet{walker2022effects}. This corresponds to fixing $\newtheta = \newpsi = \pi/2$ in the slow variables, under which the remaining emergent equation for in-plane orientation $\newphi$ \eqref{eq: slow evolution newphi} reduces significantly to
\begin{align}
\label{eq: phi reduced 2D}
\dbyd{\newphi}{\tl} = \dfrac{1}{2} \left(1 - \Breth J_0(2\Amp) \cos 2 \newphi \right).
\end{align}
Therefore, restricting motion to the 2D shear plane means that the active spheroid behaves as a passive spheroid with effective Bretherton parameter $\Breth J_0(2\Amp)$, in agreement with the 2D results of \citet{walker2022effects}. Specifically, the emergent asymmetry generated in the full 3D emergent dynamics \eqref{eq: slow evolution} vanishes in the constrained 2D dynamics. Hence, we may conclude that the emergent asymmetry that arises is a 3D effect generated by out-of-shear-plane interactions between the swimmer and the shear flow.

A natural question to ask is why symmetries appear to be preserved in the emergent dynamics arising from some types of self-generated motion. Intuitively, it seems as though an important factor should be the average shape of a rapidly moving object, with any symmetries therein conserved in the emergent dynamics. For a spheroid rapidly rotating at a constant rate about an axis fixed in the swimmer frame, the average shape is axisymmetric and the emergent dynamics are equivalent to those of a passive axisymmetric object \citep{dalwadi2024generalisedparti}. For the rapidly yawing spheroid considered here, the average shape is not axisymmetric in general. However, the average shape does have two planes of symmetry, and this symmetry is retained in the effective passive shape represented by the emergent dynamics we derive here. Curiously, however, the average shape does not tell the full story; it is not the shape of the equivalent passive object in general. This can be demonstrated by considering the aspect ratio \eqref{eq: effective aspect ratio} of the effective passive spheroid that arises when $J_0(2\Amp) = 0$ (including the large-$\Amp$ limit). In this scenario, the effective aspect ratio \eqref{eq: effective aspect ratio} is bounded between $(1/\sqrt{3},\sqrt{3})$, no matter how large or small the physical aspect ratio $\ratio$, and is therefore different from the average shape in general. While this may seem surprising due to the linearity of the Stokes equations, the difference occurs because the Stokes equations are not linear in geometry. The general nature of the relationship between the average shape of the fast motion and any effective hydrodynamic shape therefore remains an open question.

Given the technical nature of our analysis, it is instructive to consider further the averages \eqref{eq: solv conditions orig} required to determine the emergent equations. These averages are weighted nontrivially in a manner that is systematically determined by solving the (non-self) adjoint of the nonautonomous first-correction system \eqref{eq: adjoint system}, \eqref{eq: adjoint matrix} in Appendix \ref{sec: Solving adjoint system}. Requiring a technical analysis to determine the appropriate averages to take is not unusual in nonlinear multiple scales problems; since $\av{ab} \neq \av{a} \av{b}$ in general, intuitive arguments that do not properly account for nonlinearities may break down, and a key question in such problems is often \emph{which} average one should take. In fact, the averages that arise from our analysis are more straightforward to interpret if we write the original functions $\fgen$ from \eqref{eq: f functions} (which represent the slow interaction with the far-field shear flow) in terms of the angular velocity components of the spheroid in the swimmer-fixed frame $(\angvela, \angvelb, \angvelc)$:
\begin{align}
\label{eq: f functions in terms of ang vel}
\fb = \angvelb \cos \psi - \angvelc \sin \psi, \quad 
\fc = \angvela - \fa \cos \theta, \quad
\fa = \dfrac{\angvelb \sin \psi + \angvelc \cos \psi}{\sin \theta}.
\end{align}
Substituting \eqref{eq: f functions in terms of ang vel} into the right-hand sides of the averages \eqref{eq: solv conditions orig}, we see that the averages are linear combinations of the averaged quantities
\begin{align}
\label{eq: new averaged quantities}
\av{\angvela}, \quad
\av{\angvelb \cos \theta - \angvelc \sin \theta \cos \psi}, \quad
\av{\angvelb \sin \theta \cos \psi + \angvelc \cos \theta}.
\end{align}
Therefore, \eqref{eq: new averaged quantities} provides a more physical interpretation of the averages we have systematically derived via our technical analysis; the appropriate averages are specific combinations of the angular velocity components of the spheroid. Specifically, the component along the symmetry axis ($\angvela$) is averaged without modification, but the components perpendicular to this axis ($\angvelb$ and $\angvelc$) must be weighted in a manner that accounts for the plane of yawing.

We also derived the emergent translational dynamics \eqref{eq: slow ev trans} that arise from the combination of rapid yawing with self-generated translation of the spheroid centre of mass. We specifically consider self-generated motion that is oscillatory in a swimmer-fixed frame, with the same period as the yawing. Importantly, the effective translational velocity $\Veleff$ we derive in \eqref{eq: Veleff} is constant in the average orientation basis vectors $\tilde{\vec{e}}_i(\newtheta,\newpsi,\newphi)$ fixed in the average swimmer frame. We note that the effective velocity in the direction normal to the yawing plane $(\Veleff \bcdot \tilde{\vec{e}}_2)$ is simply the average of the full translational velocity \eqref{eq: Vel def fast} in this direction. Therefore, the emergent translation in the direction normal to the yawing plane is essentially unaffected by yawing (as expected intuitively) and, being independent of $\velosc_2$ and $\velps_2$ (the amplitude and phase shift of the oscillation in the direction normal to the yawing plane), also ignores any oscillation of the translational velocity in this direction. However, these properties do not carry over to the effective velocity in the yawing plane.

In the yawing plane, the emergent translational velocity has two key contributions. The first is due to the average of the full translational velocity in this plane, weighted by $J_0(\Amp)$. Physically, this is due to the yawing causing the constant translation to generate a curved trajectory in physical space, reducing the magnitude of the net translation over a yawing period. The second is due to the interaction of the yawing motion with the oscillatory part of the full translational velocity in the yawing plane. Importantly, this contribution can only arise if the translational oscillations in the yawing plane are not in phase or antiphase with the yawing oscillation (i.e. it requires the phase shifts in the yawing plane $\velps_1, \velps_3 \neq 0, \pi$), as indicated by the dependence of the effective velocity $\Veleff$ on $\sin \velps_1$ and $\sin \velps_3$ \eqref{eq: Veleff}. We also note that translational oscillations in one direction of the yawing plane generate orthogonal contributions within the yawing plane, and these contributions are weighted by $J_1(\Amp)$. Physically, this is because yawing moves the swimmer-fixed basis in the yawing plane orthogonally within the laboratory frame, and so phase differences of translational oscillations in the swimmer frame can lead to orthogonal emergent contributions.

Since we have demonstrated that rapid yawing causes a spheroid to demonstrate rotational dynamics equivalent to those for an effective object with two planes of symmetry, and objects with this type of symmetry can have translation-rotation coupling, it is interesting to note that this coupling does not arise in the emergent translational dynamics here. That is, the rapid yawing by itself does not generate any emergent translation in \eqref{eq: slow ev trans}. However, if the spheroid self-generates oscillatory self-directed motion, then this can combine with the rapid yawing to generate orthogonal effective translation. We note that this effect is independent of the shear flow in the sense that it would still occur for a self-propelling spheroid in a fluid with no externally imposed flow, though it is implicitly affected by the shear flow via the interacting flow effect on the slow evolution of $(\newtheta,\newpsi,\newphi)$.

Finally, we note that our results are straightforward to generalize to several other scenarios. Since the translational results do not affect the leading-order rotational dynamics, it is straightforward to incorporate different types of self-translation into our results by calculating the resulting average in \eqref{eq: slow ev trans midway}. For the rotational dynamics, the simplest generalization is that our results hold immediately for rapidly yawing general axisymmetric objects, now interpreting the Bretherton parameter as the measure of an effective physical aspect ratio \citep{bretherton1962motion,brenner1964stokes}. Our results can also be extended to consider general periodic yawing functions, essentially replacing $\Om \Amp \cos \left(\Om \tstandard\right)$ in \eqref{eq: yawing motion} with a general periodic function $\Om f'(\Om \tstandard)$. In this case, all our analysis up to and including \eqref{eq: solv conditions orig RHS} still holds, and the corresponding versions of the slow evolution equations \eqref{eq: slow evolution} can be obtained by simple evaluation of $\av{s^2_f}$, $\av{c^2_f}$, and $\av{s_f c_f}$ in terms of the integrated function $f(\ts)$ (imposing $f(0) = 0$, and where $\ts = \Om \tstandard$).

For odd yawing functions we have $\av{s_f c_f} = 0$, and the appropriate emergent slow evolution equations are \eqref{eq: slow evolution}, replacing $J_0(2 \Amp)$ with $\av{e^{2 \mathrm{i} f}}$. If we additionally have $\av{e^{2 \mathrm{i} f}} = 0$ then we recover the non-trivial case (3) above; the effective shape again reduces to a spheroid with axes $\widehat{a}$, $\widehat{b}$, $\widehat{b}$ and aspect ratio \eqref{eq: effective aspect ratio}. In this scenario, the nonlinear transformations
\begin{align}
\label{eq: nonlinear transform into Jeffery}
s_{\newtheta} s_{\newpsi} \mapsto c_{\newtheta}, \quad
s_{\newtheta} c_{\newpsi} \mapsto s_{\newtheta} s_{\newpsi}, \quad
c_{\newtheta} s_{\newpsi} s_{\newphi} - c_{\newpsi} c_{\newphi} \mapsto s_{\newtheta} s_{\newphi},
\end{align}
recover Jeffery's equations directly. That is, they transform \eqref{eq: slow evolution} into \eqref{eq: slow evolution outline} with $\Beff_1 = - \Beff_2 = -\Breth/2$ and $\Beff_3 = 0$. This observation provides a potential interpretation for the generation of asymmetry. Namely, considering $e^{\mathrm{i} f} = \xi+\mathrm{i}\eta$ on the complex unit circle, $\av{e^{2 \mathrm{i} f}} = 0$ corresponds to $\av{\xi^2} = \av{\eta^2}$ and $\av{\xi \eta} = 0$ i.e. the mean square orientation having no preferred direction. Hence, emergent asymmetry can arise when there is some bias in the preferred mean square orientation of rapid motion.

\textbf{Acknowledgements.} The author thanks EA Gaffney, K Ishimoto, C Moreau, and BJ Walker for helpful discussions.

\textbf{Funding.} For the purpose of Open Access, the author has applied a CC BY public copyright licence to any Author Accepted Manuscript (AAM) version arising from this submission.

\textbf{Conflict of interests.} The author reports no conflict of interest.
%

\appendix

\section{Solving the leading-order system \eqref{eq: full gov eq trans LO General}}

\label{sec: Solving LO system}

In this Appendix we solve the nonlinear, nonautonomous leading-order system \eqref{eq: full gov eq trans LO General}. We start by noting that the first two equations \eqref{eq: full gov eq trans LO General}a,b decouple from the third \eqref{eq: full gov eq trans LO General}c, and exhibit fast-time conserved quantities. To see this, we divide the second equation by the first to obtain
\begin{align}
\label{eq: dpsi by dtheta}
\pbyp{\psi_0}{\theta_0} =  - \cot \theta_0 \tan \psi_0.
\end{align} 
Then, we can integrate \eqref{eq: dpsi by dtheta} directly to deduce that
\begin{align}
\label{eq: const of integ alp}
\sin \theta_0 \sin \psi_0 = \alp(\tl),
\end{align} 
where $\alp(\tl)$ is a (slow-time) function of integration i.e. a fast-time conserved quantity.

We then substitute \eqref{eq: const of integ alp} into \eqref{eq: full gov eq trans LO General}c to obtain the governing equation
\begin{align}
\pbyp{\theta_0}{\ts}  = \Amp \cos \ts \dfrac{\sqrt{\sin^2 \theta_0 - \alp^2(\tl)}}{\sin \theta_0},
\end{align}
which can be rearranged to obtain
\begin{align}
\label{eq: sin theta integral}
\int \dfrac{\sin \theta_0 \, \mathrm{d}\theta_0}{\sqrt{1 - \alp^2(\tl) - \cos^2 \theta_0}} = \int \Amp \cos \ts \, \mathrm{d}\ts = \Amp \sin \ts + \muc(\tl),
\end{align}
where $\muc(\tl)$ is the second of three functions of integration. The integral on the left-hand side of \eqref{eq: sin theta integral} can be calculated by direct substitution of $\cos \theta_0 = \sqrt{1 - \alp^2} \cos u$, yielding the solution
\begin{align}
\label{eq: cos theta sol app}
\cos \theta_0 = \sqrt{1 - \alp^2(\tl)} \cos(\Amp \sin \ts + \muc(\tl)).
\end{align}

Finally, we can solve the remaining leading-order equation \eqref{eq: full gov eq trans LO General}c by substituting \eqref{eq: const of integ alp} and \eqref{eq: cos theta sol app} into \eqref{eq: full gov eq trans LO General}c to obtain
\begin{align}
\label{eq: phi ODE transform}
\pbyp{\phi_0}{\ts} = \dfrac{\alp(\tl) \Amp \cos \ts}{1 - (1 - \alp^2(\tl)) \cos^2(\Amp \sin \ts + \muc(\tl))}.
\end{align}
We can integrate \eqref{eq: phi ODE transform} directly using the substitution $\tan (\Amp \sin \ts + \muc) = \alp \tan u$, resulting in the solution
\begin{align}
\label{eq: tan phi sol app}
\alp(\tl) \tan(\phi_0 - \phic(\tl)) =  \tan (\Amp \sin \ts + \muc(\tl)),
\end{align}
where $\phic(\tl)$ is the third and final function of integration from the leading-order solution.

The leading-order solutions \eqref{eq: const of integ alp}, \eqref{eq: cos theta sol app}, and \eqref{eq: tan phi sol app} are the general solutions to the nonlinear, nonautonomous leading-order system \eqref{eq: full gov eq trans LO General}. These are equivalent to \eqref{eq: mu def mod} after appropriately redefining the functions of integration $(\alp(\tl), \muc(\tl), \phic(\tl))$ into $(\newtheta(\tl), \newpsi(\tl), \newphi(\tl))$. We redefine these fast-time conserved quantities so that the new functions of integration are equivalent to $(\theta, \psi, \phi)$ in the limit $\Amp \to 0$, in which case we expect to recover the original Jeffery's equations (\eqref{eq: full gov eq} with $\Amp = 0$) with no fast-time variation. Specifically, we use the following transformations:
\begin{subequations}
\label{eq: functions of integration transforms}
\begin{align}
\alp(\tl) &= \sin \newtheta(\tl) \sin \newpsi(\tl), \\ 
\tan \muc(\tl) &= \tan \newtheta(\tl) \cos \newpsi(\tl), \\ 
\tan \phic(\tl) &= \dfrac{\cos \newtheta(\tl) \tan \newpsi(\tl) \tan \newphi(\tl) - 1}{\cos \newtheta(\tl) \tan \newpsi(\tl) + \tan \newphi(\tl)}.
\end{align}
\end{subequations}

\section{Solving the adjoint system \eqref{eq: adjoint system}, \eqref{eq: adjoint matrix}}

\label{sec: Solving adjoint system}

In this Appendix we solve the linear, nonautonomous adjoint system \eqref{eq: adjoint system}, \eqref{eq: adjoint matrix} for $\bsX = (\bsXa, \bsXb, \bsXc) $ with periodic boundary conditions. This solution will allow us to generate the appropriate solvability conditions, and hence emergent equations, via \eqref{eq: general solv cond}. We note that the solution method we present here is a modified version of that in \citet{dalwadi2024generalisedparti} for a similar (but autonomous) adjoint system.

Since we have a three-dimensional linear system, we seek three linearly independent solutions. We start by considering the bottom row of $L^*$ in \eqref{eq: adjoint matrix}, which tells us that $\bsXc = \text{constant}$ in all solutions. We can reduce our task to solving a two-dimensional system by setting this constant equal to zero in two of the linearly independent solutions, ensuring that the third solution will be able to generate our solution basis for $\bsXc$. For this third solution, we explicitly write
\begin{align}
\label{eq: Xc sol}
\bsXc = \Cc,
\end{align}
where $\Cc$ is an arbitrary constant. Substituting \eqref{eq: Xc sol} into \eqref{eq: adjoint system}, \eqref{eq: adjoint matrix}, we obtain
\begin{subequations}
\label{eq: 2D adjoint gov}
\begin{align}
\label{eq: \bsXa gov}
\dbyd{\bsXa}{\ts} &= \dfrac{\Amp \cos \ts  \sin \psi_0}{\sin^2 \theta_0} \left[- \bsXb + \Cc \cos \theta_0  \right], \\
\label{eq: \bsXb gov}
\dbyd{\bsXb}{\ts} &= \Amp \cos \ts \left[\bsXa \sin \psi_0 + \dfrac{\cos \psi_0}{\sin \theta_0} \left(\cos \theta_0 \bsXb - \Cc \right) \right].
\end{align}
\end{subequations}
Given that the right-hand side of \eqref{eq: \bsXa gov} does not depend on $\bsXa$, it is convenient to introduce $\bsXb = \Cc \cos \theta_0 + \tilde{\bsXb}$ to reduce the complexity of the system. In this case, the time derivative generated by differentiating $\cos \theta_0$ in \eqref{eq: \bsXb gov} cancels exactly with the inhomogeneous term on the right-hand side, and so \eqref{eq: 2D adjoint gov} becomes
\begin{subequations}
\label{eq: 2D adjoint gov after C3}
\begin{align}
\label{eq: \bsXa gov after C3}
\dbyd{\bsXa}{\ts} &= -\dfrac{\Amp \cos \ts  \sin \psi_0 }{\sin^2 \theta_0}  \tilde{\bsXb}, \\
\label{eq: \bsXb gov after C3}
\dbyd{\tilde{\bsXb}}{\ts} &= \Amp \cos \ts \left[\bsXa \sin \psi_0 +  \dfrac{\cos \psi_0 \cos \theta_0}{\sin \theta_0} \tilde{\bsXb} \right].
\end{align}
\end{subequations}
That is, we  have transformed into a homogeneous system, which is solved by $(\bsXa, \tilde{\bsXb}) = (0, 0)$. Hence, we have generated one nontrivial solution to \eqref{eq: adjoint system}, \eqref{eq: adjoint matrix}, namely
\begin{align}
\label{eq: LI adjoint sol 3}
\bs{Y} = \Cc  (0, \cos \theta_0, 1)^{\intercal}.
\end{align}

As noted above, the remaining two solutions can be determined by now setting $\Cc = 0$ in \eqref{eq: 2D adjoint gov} to obtain the system 
\begin{subequations}
\label{eq: 2D adjoint gov no C}
\begin{align}
\label{eq: \bsXa gov no C}
\dbyd{\bsXa}{\ts} &= -\dfrac{\Amp \cos \ts  \sin \psi_0}{\sin^2 \theta_0}\bsXb, \\
\label{eq: \bsXb gov no C}
\dbyd{\bsXb}{\ts} &= \Amp \cos \ts \left[\bsXa \sin \psi_0 + \dfrac{\cos \psi_0 \cos \theta_0}{\sin \theta_0} \bsXb \right].
\end{align}
\end{subequations}

To derive two linearly independent solutions to \eqref{eq: 2D adjoint gov no C}, we seek nonlinear transformations to map the remaining adjoint problem \eqref{eq: 2D adjoint gov no C} into a homogeneous version of the first-correction system \eqref{eq: full gov eq trans oma O1 eps}. The reason this is helpful is because the linear operator of \eqref{eq: full gov eq trans oma O1 eps} is a perturbed version of the (nonlinear) operator of the leading-order system \eqref{eq: full gov eq trans LO General}. Hence, perturbed versions of the solutions we derived to the leading-order system \eqref{eq: mu def mod} will solve the homogeneous version of the first-correction system \eqref{eq: full gov eq trans oma O1 eps}. That is,
\begin{align}
\label{eq: homog next order sol}
\theta_1 = \dfrac{1}{\Amp \cos \ts}\pbyp{\theta_0}{\ts} = \cos \psi_0, \quad \psi_1 = \dfrac{1}{\Amp \cos \ts}\pbyp{\psi_0}{\ts} = - \dfrac{\cos \theta_0 \sin \psi_0}{\sin \theta_0},
\end{align}
solve the homogeneous version of the first-correction system \eqref{eq: theta eq trans oma O1 eps},  \eqref{eq: psi eq trans oma O1 eps}.

To determine how to map \eqref{eq: 2D adjoint gov no C} into a homogeneous version of \eqref{eq: full gov eq trans oma O1 eps}, we introduce the transformations
\begin{align}
\label{eq: mappings}
\bsXa(\ts) = \zeta_1(\theta_0) \tilde{\bsXa}(\ts), \quad \bsXb(\ts) = \zeta_2(\theta_0) \tilde{\bsXb}(\ts),
\end{align}
where the $\zeta_i$ are the (as-of-yet) unknown nonlinear functions of $\theta_0$ we seek. Substituting \eqref{eq: mappings} into \eqref{eq: 2D adjoint gov no C}, we obtain
\begin{subequations}
\label{eq: 2D adjoint gov no C transform}
\begin{align}
\label{eq: \bsXa gov no C transform}
\dbyd{\tilde{\bsXa}}{\ts} &= -\Amp \cos \ts  \left[\dfrac{\zeta'_1}{\zeta_1} \cos \psi_0 \tilde{\bsXa} + \dfrac{\zeta_2}{\zeta_1} \dfrac{ \sin \psi_0}{\sin^2 \theta_0} \tilde{\bsXb} \right], \\
\label{eq: \bsXb gov no C transform}
\dbyd{\tilde{\bsXb}}{\ts} &= \Amp \cos \ts \left[\dfrac{\zeta_1}{\zeta_2} \sin \psi_0 \tilde{\bsXa} + \cos \psi_0 \left( \dfrac{ \cos \theta_0}{\sin \theta_0} - \dfrac{\zeta'_2}{\zeta_2}\right) \bsXb \right].
\end{align}
\end{subequations}
Then, we compare \eqref{eq: 2D adjoint gov no C transform} to the homogeneous version of the first-correction system \eqref{eq: theta eq trans oma O1 eps}, \eqref{eq: psi eq trans oma O1 eps}. We see that $(\tilde{\bsXa}, \tilde{\bsXb}) = (\theta_1, \psi_1)$ when
\begin{align}
\dfrac{\zeta'_1}{\zeta_1} = 0, \quad
\dfrac{\zeta'_2}{\zeta_2} = \dfrac{2 \cos \theta_0}{\sin \theta_0}, \quad
\dfrac{\zeta_2}{\zeta_1} = \sin^2 \theta_0.
\end{align}
That is, when
\begin{align}
\label{eq: zeta transform}
\zeta_1 = \Cb, \quad
\zeta_2 = \Cb \sin^2 \theta_0,
\end{align}
where $\Cb$ is an arbitrary constant. Then substituting $(\tilde{\bsXa}, \tilde{\bsXb}) = (\theta_1, \psi_1)$ into \eqref{eq: mappings} and using the results \eqref{eq: homog next order sol} and \eqref{eq: zeta transform} gives us the second linearly independent solution
\begin{align}
\label{eq: LI adjoint sol 2}
\bs{Y} = \Cb  (\cos \psi_0, -\sin \theta_0 \cos \theta_0 \sin \psi_0, 0)^{\intercal}.
\end{align}

The third and final linearly independent solution to the adjoint problem \eqref{eq: adjoint system}, \eqref{eq: adjoint matrix} arises from comparing \eqref{eq: 2D adjoint gov no C transform} to the homogeneous version of the first-correction system \eqref{eq: theta eq trans oma O1 eps}, \eqref{eq: psi eq trans oma O1 eps} and now looking for $(\tilde{\bsXa}, \tilde{\bsXb}) = (\psi_1, \theta_1)$. This requires
\begin{align}
\dfrac{\zeta'_1}{\zeta_1} = 0, \quad
\dfrac{\zeta'_2}{\zeta_2} = 2 \dfrac{\cos \theta_0}{\sin \theta_0}, \quad
\dfrac{\zeta_2}{\zeta_1} = \sin^2 \theta_0,
\end{align}
which we can solve straightforwardly to obtain
\begin{align}
\label{eq: zeta transform second}
\zeta_1 = \Ca \sin \theta_0, \quad
\zeta_2 = -\Ca \sin \theta_0.
\end{align}
Finally, substituting $(\tilde{\bsXa}, \tilde{\bsXb}) = (\psi_1, \theta_1)$ into \eqref{eq: mappings} and using the results \eqref{eq: homog next order sol} and \eqref{eq: zeta transform second} gives us the final linearly independent solution
\begin{align}
\label{eq: LI adjoint sol 1}
\bs{Y} = \Ca  (\sin \theta_0 \cos \psi_0, \cos \theta_0 \sin \psi_0, 0)^{\intercal}.
\end{align}
Together, the three linearly independent solutions \eqref{eq: LI adjoint sol 3}, \eqref{eq: LI adjoint sol 2}, \eqref{eq: LI adjoint sol 1} to the adjoint problem \eqref{eq: adjoint system}, \eqref{eq: adjoint matrix} give the general solution \eqref{eq: adjoint solution}.

\bibliographystyle{jfm_draft}
\bibliography{reference.bib}

\end{document}